\documentclass[amsmath,amssymb,aps,prx,reprint,superscriptaddress,footnoteinbib,longbibliography]{revtex4-1}
\usepackage[utf8]{inputenc}
\usepackage[T1]{fontenc}
\usepackage{amsmath,amsfonts,amssymb,mathtools,dsfont}
\usepackage{graphicx}
\usepackage[hidelinks]{hyperref}
\usepackage{url}

\graphicspath{{./figures/}}

\begin{document}

\title{Controlling rotations of magnetically levitated superconductors}

\author{Fynn Köller}
\affiliation{Institute for Complex Quantum Systems and Center for Integrated Quantum Science and Technology, Ulm University, Albert-Einstein-Allee 11, 89069 Ulm, Germany}

\author{Dominik Maile}
\affiliation{Institute for Complex Quantum Systems and Center for Integrated Quantum Science and Technology, Ulm University, Albert-Einstein-Allee 11, 89069 Ulm, Germany}

\author{Joachim Ankerhold}
\affiliation{Institute for Complex Quantum Systems and Center for Integrated Quantum Science and Technology, Ulm University, Albert-Einstein-Allee 11, 89069 Ulm, Germany}

\author{Klaus Hornberger}
\affiliation{University of Duisburg-Essen, Faculty of Physics, Lotharstra\ss e 1, 47057 Duisburg, Germany}

\author{Witlef Wieczorek}
\affiliation{Department of Microtechnology and Nanoscience (MC2),
Chalmers University of Technology, SE-412 96 Göteborg, Sweden}

\author{Benjamin A. Stickler}
\affiliation{Institute for Complex Quantum Systems and Center for Integrated Quantum Science and Technology, Ulm University, Albert-Einstein-Allee 11, 89069 Ulm, Germany}

\begin{abstract}
Magnetically trapped type-I superconductors are promising candidates for precision experiments at the quantum-to-classical borderline, with applications in quantum sensing of forces and accelerations and fundamental tests of quantum physics. Here, we show how the rotational motion of micron-sized superconductors is strongly affected by  (i) the diamagnetic torques due to the gradient of the trapping field and by (ii) the gyromagnetic coupling due to Einstein-de Haas and Barnett effects. We show that this allows the three-dimensional alignment of asperhical superconductors in the trap center, as required for future sensing applications and quantum experiments, and determine the resulting librational trapping frequencies. Finally, we propose an experiment to probe gyromagnetic coupling in levitated superconductors and we discuss how it can be used to control the particle rotation.
\end{abstract}

\maketitle

\section{Introduction}

A type-I superconducting particle in a magneto-static field experiences a force opposing the field, enabling its stable levitation in the minimum of a magnetic quadrupole trap \cite{geim2000}. Compared to electric ion traps or optical tweezers, magnetic platforms typically come with small trapping frequencies and low noise levels. This makes these systems highly attractive for precision sensing \cite{goodkind1999superconducting,pratcamps2017,timberlake2019,vinante2020,lewandowski2021} and quantum superposition tests with massive particles \cite{cirio2012,romeroisart2012,rusconi2017,pino2018,rusconi2022,kustura2022,raman2025,cunill2026macroscopic}. State-of-the-art experiments magnetically levitate micron-sized superconductors and demonstrate that their center-of-mass motion can be precisely monitored and controlled \cite{latorre2020,navau2021,latorre2022,latorre2023,hofer2022high,Schmidt2024,smit2026,hansen2026}. These experiments present a first step towards cooling the particle motion to low temperature, perhaps even into the quantum regime \cite{hofer2022high,Schmidt2024}, as required for future sensing applications and experiments probing the coupling between co-levitated particles.  

The rotational motion of a levitated type-I superconductor can be strongly affected by the trapping field due to two distinct mechanisms: (i) the induced magnetization couples to the external trapping fields, giving rise to a diamagnetic torque \cite{jackson1999}, and (ii) the induced magnetization contributes to the total angular momentum of the particle, inducing what is known as gyromagnetic coupling \cite{barnett1935gyro,vanvleck1951thecoupling,rusconi2016magnetic,rusconi2017,ma2021torque,rusconi2022,kustura2022,wachter2025gyroscopically,ahrens2026observation}. The diamagnetic torque vanishes for objects of perfectly spherical shape, while gyromagnetic coupling prevails even for spherical particles. However, since perfect isotropy is never achieved experimentally, both effects are relevant for future precision experiments with levitated superconductors.

In this work, we derive the diamagnetic force and torque and the gyromagnetic coupling for superconducting ellipsoids suspended in a magnetic quadrupole field. While the force and gyromagnetic coupling are fully determined by the induced dipole vector, determining the diamagnetic torque also requires accounting for the induced quadrupole moments. These induced moments depend on the local magnetic field strength and its gradient as well as on the relative orientation between the local field lines and the principal axes of the particle, leading to a coupling between its rotation and center-of-mass motion. It is well established that the induced dipole moment can be calculated analytically for ellipsoidal objects in terms of the so-called demagnetization factors \cite{stoner1945xcvii,osborn1945demagnetizing,bortsoldevilla2024}, providing excellent analytical approximations even for particles of non-ellipsoidal shape \cite{hofer2024,bortsoldevilla2024}. The core theoretical innovation of this work is to derive analytical expressions for the induced quadrupole tensor by generalizing a method devised by Dirichlet that can be used to calculate the induced dipole moment \cite{dittrich2016dirichlet}. This allows us to obtain closed-form analytical expressions for the diamagnetic forces and torques, including gyromagnetic coupling.

We use these findings to show how aspherical superconductors can be three-dimensionally aligned in the trap center, to calculate the resulting center-of-mass and librational trapping frequencies, and to propose an experiment probing gyromagnetic coupling in levitated superconductors. We expect our results to be instrumental for future precision experiments with  levitated superconductors in the quantum regime, as required for sensing applications \cite{gonzalez2021,stickler2021}, for probing dark-matter models \cite{moore2021,Higgins2024}, for gravimetry \cite{griggs2017,fuchs2024,Carney2025,headly2026quantum,amaral2026}, and perhaps even for future tests of quantum aspects of Newtonian gravity \cite{bose2017,marletto2017,lami2024testing,higgins2024superrot,bulling2026stability}. Moreover, the developed theoretical description can be used to calculate the induced magnetization fields in magnetizable particles in general, including diamagnetic as well as paramagnetic objects. Our results can thus also be expected to become relevant for ongoing experiments with other magnetizable materials, such as levitated diamonds \cite{perdriat2021spin}.

The article is structured as follows: In Sec.~\ref{sec:magrot} we review the description of rigid-body rotations of magnetic objects, Sec.~\ref{sec:mainres} summarizes our main results for levitated superconductors, whose detailed derivation is presented in Secs.~\ref{sec:derivationinducedmoments} and \ref{sec:dynamics}, including the impact of Earth's gravity. Finally, we conclude in Sec.~\ref{sec:outlook}.

\section{Dynamics of magnetic rotors}\label{sec:magrot}

Before discussing the dynamics of levitated superconductors, we first review the dynamics of magnetic rigid bodies in magnetostatic quadrupole fields. For this sake, we consider an ellipsoidally shaped rigid magnet located at the center-of-mass position ${\bf R}$ with orientation $\Omega$. The latter can be parametrized by Euler angles $\Omega = (\alpha,\beta,\gamma)$ in the $z$-$y'$-$z''$ convention, so that the ellipsoid's principal axes are ${\bf n}_i(\Omega) = {\rm R}(\Omega){\bf e}_i$, with the orthogonal rotation tensor ${\rm R}(\Omega) = {\rm R}_z(\alpha){\rm R}_y(\beta) {\rm R}_z(\gamma)$ and the space-fixed coordinate frame ${\bf e}_i$, $i = 1,2,3$ \cite{goldstein2002classical}. The particle is subject to the external field
\begin{equation}\label{eq:Bext}
{\bf B}_{\rm ext}({\bf r}) = {\bf B}_h + {\rm C}{\bf r},
\end{equation}
the sum of a homogeneous field  ${\bf B}_h$ and a quadrupole field, determined by the symmetric and traceless tensor ${\rm C}=\nabla\otimes\mathbf{B}_{\rm ext}(\mathbf{r})$ describing the field gradient.

The magnet possesses an internal magnetization field ${\bf M}({\bf r})$, implying that the external field \eqref{eq:Bext} gives rise to a force and a torque acting on the body. The force and torque on the body follow from the Lorentz force density in the external field \eqref{eq:Bext}, ${\bf f}_{\rm L}({\bf r}) = [\nabla \times {\bf M}({\bf r})] \times {\bf B}_{\rm ext}({\bf r})$, which is integrated over the position and orientation-dependent particle volume $V({\bf R},\Omega)$, see App.~\ref{sec:eom}.  Denoting the particle mass by $M$ the center-of-mass equation of motion is,
\begin{subequations}\label{eq:eom}
\begin{equation}
    \frac{d}{dt} \left ( M \dot{\bf R} \right ) =  {\rm C} {\bf m}.
\end{equation}
The total force acting on the body is thus proportional to the field gradient and to the magnetic dipole vector,
\begin{equation}\label{eq:magdipdef}
    {\bf m} = \int_{V} d^3 {\bf r} \, {\bf M}({\bf r}).
\end{equation}
When determining the rotational equations of motion care has to be taken because the body's magnetic moment contributes to its total angular momentum due to the Einstein de Haas and Barnett effects, see Sec.~\ref{sec:edhbarnett}.  Denoting the inertia tensor by ${\rm I}(\Omega)$, with inertia moments $I_i = M (a_j^2 + a_k^2)/5$ with $i\neq j \neq k$, we obtain the rotational dynamics as
\begin{equation}\label{eq:magnetrotdyn}
    \frac{d}{dt} \left ( {\rm I} \boldsymbol{\omega} - \frac{1}{\gamma_0} {\bf m} \right ) = {\bf m} \times {\bf B}_{\rm ext} + \frac{1}{3} \sum_{i = 1}^3 C_i ({\rm Q}{\bf e}_i)\times {\bf e}_i.
\end{equation}
Here, $\gamma_0$ is the gyromagnetic ratio and we chose the space-fixed coordinate frame ${\bf e}_i$, $i = 1,2,3$ such that it diagonalizes the field gradient tensor ${\rm C}{\bf e}_i = C_i {\bf e}_i$ with $C_1 + C_2 + C_3 = 0$. The change in total angular momentum is thus determined by the local torque on the dipole moment as well as by the misalignment between the field gradient and the quadrupole tensor, see App.~\ref{sec:momentsgen},
\begin{align}\label{eq:Qdef}
    {\rm Q} = &\; \int_{V} d^3{\bf r}\, \left [ \vphantom{\frac{2}{3}}3 {\bf M}({\bf r}) \otimes ({\bf r} - {\bf R}) + 3 ({\bf r} - {\bf R}) \otimes {\bf M}({\bf r})\right.\nonumber\\
    & \left. - 2 {\bf M}({\bf r}) \cdot ({\bf r} - {\bf R}) \mathds{1}\right ].
\end{align}
\end{subequations}
The latter contribution is particularly relevant close to the trap center, where ${\bf B}_{\rm ext}$ vanishes.

\begin{figure*}[t]
    \centering
    \includegraphics[width=\textwidth]{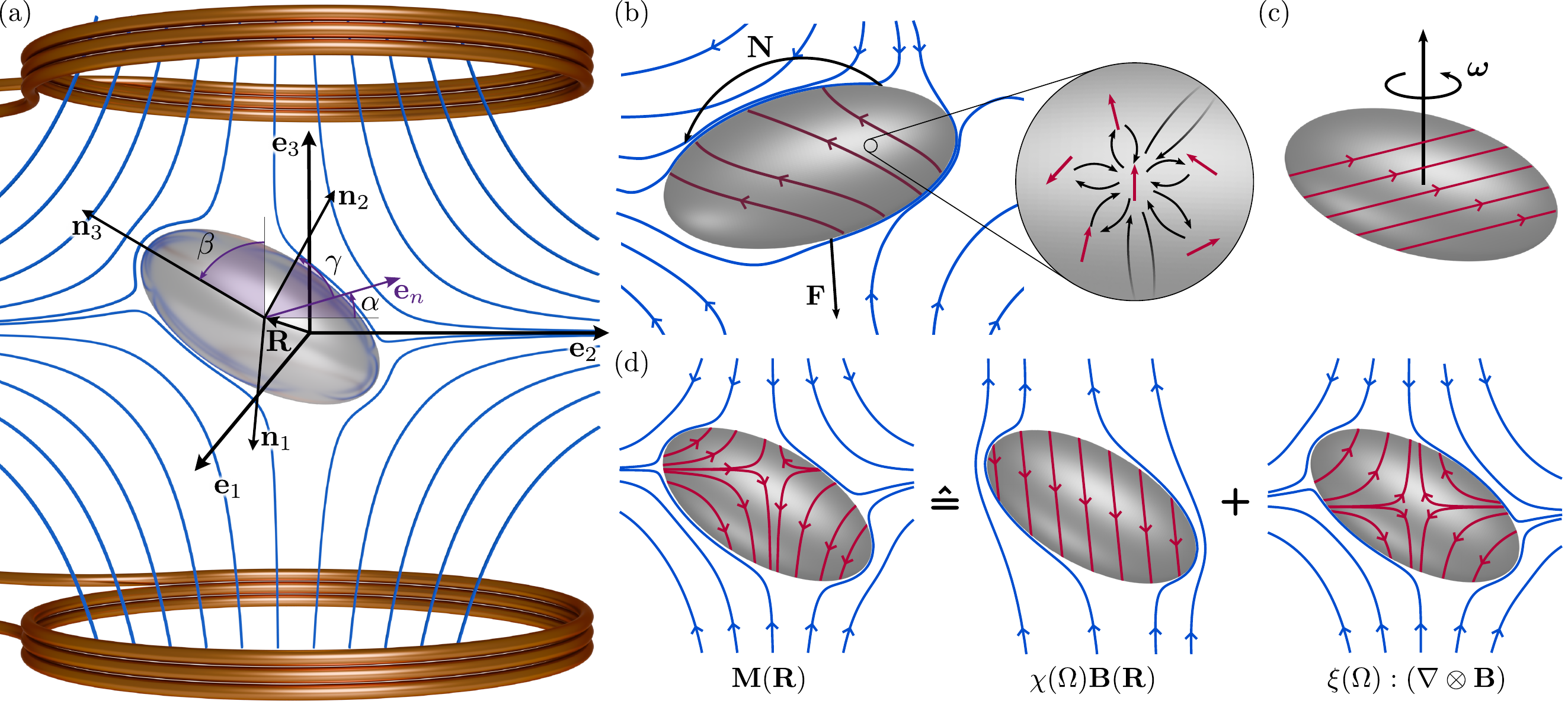}
    \caption{{\bf Rotations of levitated superconductors:} (a) Micron-sized superconductors can be stably trapped in magnetic quadrupole fields generated by two coils in anti-Helmholtz configuration. The superconductor is attracted towards the trap center due to its diamagnetic response, where its principal axes align with the axes of the trapping field. The six degrees of freedom of the particle's motion are its center-of-mass position ${\bf R}$ and orientation, expressed via the Euler-angles $\Omega = \{\alpha, \beta, \gamma\}$ in $z$-$y'$-$z''$ convention. The orientation is specified by the rotation tensor, that relates the space-fixed frame ${\bf e}_k$ with the principal-axes frame ${\bf n}_k(\Omega) = {\rm R}(\Omega) {\bf e}_k$. (b) External magnetic fields generate supercurrents on the particle surface (dark red lines), which shield the trapping field from the interior of the particle and thereby exert a magnetic force ${\bf F}$ and torque ${\bf N}$ on the superconductor. The torque near the trap center, where the field vanishes, is dominated by the magnetic quadrupole torque due to the non-vanishing field gradients. The induced supercurrents can be equivalently expressed through an internal magnetization field (red arrows in inset) that accounts for the total local field for a given particle shape and orientation. (c) Rotational motion of the superconductor also induce surface currents due to gyromagnetic coupling, also dubbed London moment. In a quadrupole trap this results in  additional forces and torques acting on the particle. (d) The internal magnetization field can be expanded into orientation-dependent and shape-dependent interior multipole moments. For quadrupole trapping fields and ellipsoidal particles, the total magnetization is exactly given by the sum of an induced dipole vector and an induced quadrupole tensor.}\label{fig:fig1}
\end{figure*}

\section{Main results}\label{sec:mainres}

We are now in a position to present our main results. We consider a micron-sized type-I superconductor of ellipsoidal shape trapped in the magnetic quadrupole field \eqref{eq:Bext}, see Fig.~\ref{fig:fig1}. In the practically relevant case where the particle extension clearly exceeds the London penetration depth, the particle behaves as a perfect diamagnet. This implies that the trapping field \eqref{eq:Bext} induces a supercurrent density ${\bf j}_s({\bf r})$ on the particle surface, which screens the magnetic field from the interior of the particle. Since there are no free charges, the current density is divergence-less and  can be expressed formally through an induced magnetization field ${\bf j}_s({\bf r}) = \nabla \times {\bf M}({\bf r})$. This induced magnetization field ${\bf M}({\bf r})$ determines the induced dipole vector \eqref{eq:magdipdef} and quadrupole tensor \eqref{eq:Qdef} that enter the magnetic force and torque Eqs.~\eqref{eq:eom} acting on the superconductor. In order to calculate these induced moments, we will now consider the more general case of a magnetizable body of homogeneous bulk permeability $\mu_r$. The limit $\mu_r \rightarrow 0$ of a perfect diamagnet can then be drawn at the end of the calculation.

\subsection{Induced magnetic moments}\label{sec:moments}

In order to calculate the induced magnetization field, we must account for the fact that the particle rotation gives rise to a synthetic magnetic field, or Barnett field. In the case of superconductors, the resulting rotation-induced magnetic moment is known as the London moment. Specifically, if the body rotates with angular velocity $\boldsymbol{\omega}$, the rotation induces the Barnett field $-\boldsymbol{\omega}/\gamma_0$, with the positive material-specific gyromagnetic ratio $\gamma_0 > 0$. The total magnetic field experienced by the body is thus given by
\begin{equation}\label{eq:quadfield}
{\bf B}_0({\bf r}) = {\bf B}_{\rm ext}({\bf r}) - \frac{\boldsymbol{\omega}}{\gamma_0},
\end{equation}
and it is this field that induces the magnetization field ${\bf M}({\bf r})$ inside the body. 

The strength and direction of the magnetization field is determined by the linear integral equation, see Sec.~\ref{sec:inducedmagn},
\begin{equation}\label{eq:intmag}
    {\bf M}({\bf r}) = (\mu_r - 1) \left [\frac{{\bf B}_{0}({\bf r})}{\mu_0} +\int_V d^3 {\bf s}\, \mathrm{G}({\bf r}-{\bf s}){\bf M}({\bf s})\right],
\end{equation}
where we integrate over the particle volume $V=V(\mathbf{R},\Omega)$ as a function of the particle position ${\bf R}$ and orientation $\Omega$. The first term on the right hand side describes the local response of the material to the effective field, while the second term accounts for the magnetic field originating from the magnetization field inside the body as described by the static dipole Green tensor ${\rm G}({\bf r}) = \nabla \otimes \nabla 1/4\pi r$. The magnetization field vanishes identically outside the ellipsoid, and it can be calculated by formally solving Eq.~\eqref{eq:intmag} as
\begin{equation}\label{eq:magnetization}
    {\bf M}({\bf r}) = \frac{1}{\mu_0} \int_V d^3{\bf s}\, {\rm K}({\bf r},{\bf s}) {\bf B}_0({\bf s}),
\end{equation}
provided the integral kernel fulfills the integral equation
\begin{align}\label{eq:inteqk}
    {\rm K}({\bf r},{\bf s}) = &\; (\mu_r - 1) \nonumber\\
    & \times \left [\delta ({\bf r} - {\bf s}) \mathds{1} + \int_V d^3{\bf s}' \, \mathrm{G}({\bf r} - {\bf s}')\mathrm{K}({\bf s}',{\bf s})\right ].
\end{align}
The latter is symmetric, ${\rm K}^T({\bf r},{\bf s}) = {\rm K}({\bf s},{\bf r})$. Equation \eqref{eq:magnetization} shows that the induced magnetization vanishes if there is no effective field. It allows calculating the induced magnetic moments for bodies of arbitrary size, shape, and permeability.

Furthermore, equation \eqref{eq:magnetization} implies that for ellipsoidally shaped bodies in an external quadrupole field only the induced dipole and quadrupole moments are non-zero, see Fig.~\ref{fig:fig1}. Both these moments can be determined by a lengthy but direct calculation presented in Sec. \ref{sec:derivationinducedmoments}. This calculation shows that the induced dipole vector is
\begin{subequations}
\begin{equation}\label{eq:magdip}
    {\bf m} = \frac{V_0}{\mu_0}\chi(\Omega) \left [ {\bf B}_{\rm ext}({\bf R}) - \frac{\boldsymbol{\omega}}{\gamma_0}\right ],
\end{equation}
where $V_0$ denotes the particle volume. The dipole vector is thus proportional to the effective field experienced by the body and transformed as described by the shape and orientation-dependent rank-two susceptibility tensor
\begin{equation}\label{eq:chi2}
    \chi(\Omega) = \sum_{i = 1}^3 \chi_i {\bf n}_i \otimes {\bf n}_i.
\end{equation}
\end{subequations}
This tensor is diagonal in the principal axis frame and fully determined by the susceptibilities $\chi_i$, see Eq.~\eqref{eq:chii}. For spherical particles the tensor turns isotropic and the dipole vector aligns with the total field \eqref{eq:quadfield}. The susceptibilities have also been derived in \cite{stoner1945xcvii,osborn1945demagnetizing,bortsoldevilla2024} and can be re-expressed in terms of Jacobi elliptic integrals. Ref.~\cite{bortsoldevilla2024} demonstrates that the dipole moment induced in an ellipsoid well approximates also that induced in non-ellipsoidal particles, such as cylinders and prisms.

The induced quadrupole tensor can be obtained in a conceptually similar calculation, yielding
\begin{subequations}
\begin{align}\label{eq:magquad}
    {\rm Q} = &\; \frac{V_0}{\mu_0}\xi(\Omega) : [\nabla \otimes {\bf B}_{\rm ext}({\bf R})] = \frac{V_0}{\mu_0}\xi(\Omega) : {\rm C}.
\end{align}
Thus the induced quadrupole tensor is proportional to the field gradient tensor ${\rm C} = \nabla \otimes {\bf B}_{\rm ext}({\bf r})$. It is obtained by contraction with the rank-four susceptibility tensor $\xi$ with the double-dot product defined by  $({\bf e}_i\otimes {\bf e}_j \otimes {\bf e}_k \otimes {\bf e}_\ell):({\bf e}_m\otimes {\bf e}_n) =  \delta_{km}\delta_{\ell n}{\bf e}_i \otimes {\bf e}_j$ and by linearity. The rank-four susceptibility tensor is tied to the principal-axis frame and depends only on the particle orientation,
\begin{align}\label{eq:xi4}
    \xi\left(\Omega\right) = &\; \sum_{i,j = 1\atop{i\neq j}}^{3} \left[\xi_{ij}^{(1)} {\bf n}_i \otimes {\bf n}_j \otimes \left({\bf n}_i \otimes {\bf n}_j + {\bf n}_j \otimes {\bf n}_i\right)\right. \nonumber\\
    &\left. + \xi_{ij}^{(2)} {\bf n}_i \otimes {\bf n}_i \otimes ({\bf n}_j \otimes {\bf n}_j- {\bf n}_i \otimes {\bf n}_i )\right ].
\end{align}
\end{subequations}
Analytic expressions for the symmetric matrix elements $\xi_{ij}^{(\ell)} = \xi_{ji}^{(\ell)}$ for $\ell = 1,2$ are provided in Sec.~\ref{sec:derivationinducedmoments}. The components of the rank-four susceptibility tensor $\xi_{ijk\ell} = ({\bf n}_i \otimes {\bf n}_j) : \xi(\Omega) : ( {\bf n}_k \otimes {\bf n}_\ell)$ are symmetric under transposition of the two leftmost and rightmost indices $\xi_{ijk\ell} = \xi_{jik\ell} = \xi_{ij\ell k}$ as well as the two inner and outer indices $\xi_{ijk\ell} = \xi_{\ell kji}$, implying that \eqref{eq:magquad} is also symmetric. The partial trace over the two leftmost indices as well as over the two rightmost indices of the rank-four susceptibility tensor vanish so that the quadrupole tensor \eqref{eq:magquad} is traceless.

\subsection{Dynamics of levitated superconductors}\label{sec:eqnmotion}

We can now use the induced moments \eqref{eq:magdip} and \eqref{eq:magquad} to obtain the equations of motion \eqref{eq:eom} for a levitated superconductor. This yields
\begin{subequations}\label{eq:eomfinal}
\begin{equation}\label{eq:eomfinl1}
\frac{d}{dt}(M \dot{\bf R}) = \frac{V_0}{\mu_0} {\rm C}\chi \left [ {\bf B}_{\rm ext}({\bf R}) - \frac{\boldsymbol{\omega}}{\gamma_0} \right ],
\end{equation}
and
\begin{align}\label{eq:eomfinal2}
\frac{d}{dt} \left [ {\rm I}_{\rm eff} \boldsymbol{\omega} - \frac{V_0}{\mu_0 \gamma_0} \chi {\bf B}_{\rm ext} \right ] 
 = & \frac{V_0}{\mu_0} \chi \left [ {\bf B}_{\rm ext} - \frac{\boldsymbol{\omega}}{\gamma_0} \right ] \times {\bf B}_{\rm ext}  \notag\\
& +\frac{V_0}{3 \mu_0}\sum_{i = 1}^3 C_i [(\xi:{\rm C}){\bf e}_i]\times {\bf e}_i
\end{align}
\end{subequations}
where ${\bf B}_{\rm ext} = {\bf B}_{\rm ext}({\bf R})$ and we define the effective inertia tensor,
\begin{equation}
    {\rm I}_{\rm eff}(\Omega) = {\rm I}(\Omega) + \frac{V_0}{\mu_0 \gamma_0^2} \chi(\Omega),
\end{equation}
renormalized due to the Einstein-de Haas and Barnett effects. For realistic particle sizes and particle shapes, this renormalization is negligible, ${\rm I}_{\rm eff} \approx {\rm I}$.

The equations of motion \eqref{eq:eomfinal} describe how the motion of the particle is affected by the supercurrents induced on its surface. These supercurrents are caused by both the local magnetic field and its gradient as well as by the particle rotation, giving rise to a gyromagnetic coupling between center-of-mass motion and rotational dynamics. Moreover, the induced quadrupole tensor causes a torque that dominates the dipole torque close to the trap center, where the trapping field vanishes. We will see below that the combination of gyromagnetic coupling and quadrupole torque allows stable trapping, alignment, and controlled rotation of the particle close to the center of the quadrupole trap.

The equations of motion \eqref{eq:eomfinal} are conservative and thus generated by a Lagrangian. As shown in Sec.~\ref{sec:lagrangian}, this Lagrangian is given by
\begin{align}\label{eq:lagrangian}
    L = &\; \frac{M}{2} \dot{\bf R}^2 + \frac{1}{2} \boldsymbol{\omega} \cdot {\rm I}_{\rm eff} \boldsymbol{\omega} - \frac{V_0}{\mu_0\gamma_0} \boldsymbol{\omega} \cdot \chi {\bf B}_{\rm ext} \nonumber\\
    & + \frac{V_0}{2\mu_0} {\bf B}_{\rm ext} \cdot \chi {\bf B}_{\rm ext} + \frac{V_0}{12\mu_0} {\rm Tr}[(\xi : {\rm C}){\rm C}],
\end{align}
where ${\rm Tr}(\cdot)$ denotes the tensor trace. The first two terms are the kinetic energy of center-of-mass motion and rotations, respectively. The third term in Eq.~\eqref{eq:lagrangian} describes the coupling between the particle rotation and its center-of-mass motion due the Einstein-de Haas and Barnett effects. The fourth and fifth terms describe the energy of the induced magnetic dipole and quadrupole moment in the applied magnetic field, respectively. The dipole potential leads to stable levitation in the trap center, where the quadrupole potential tends to align asymmetric rotors with the trapping geometry. While this quadrupole potential can be safely neglected far from the trap center (at high motional temperatures), it becomes increasingly relevant close to the trap center (at low temperatures) where the dipole torque vanishes. An explicit expression of the Lagrangian \eqref{eq:lagrangian} in terms of Euler angles $\Omega = (\alpha,\beta,\gamma)$  is given in Sec.~\ref{sec:lagrangian}.

The Lagrangian \eqref{eq:lagrangian} implies that the canonical angular momentum
\begin{equation}\label{eq:canonicalangularmomentum}
    {\bf J} = \frac{\partial L}{\partial \boldsymbol{\omega}} = {\rm I}_{\rm eff} \boldsymbol{\omega} - \frac{V_0}{\mu_0 \gamma_0}\chi {\bf B}_{\rm ext}.
\end{equation}
is not the kinetic angular momentum but the total angular momentum of the body, containing the contribution of the supercurrents. The latter enter through the effective inertia as well as through the induced magnetic dipole moment, as described by the second term. This gyromagnetic contribution expresses the fact that magnetization, and thus supercurrents on the particle surface, get dynamically converted into mechanical rotation, as described in the Einstein-de Haas effect. At the same time, the canonical linear momentum
\begin{equation}\label{eq:canonicallinearmomentum}
    {\bf P} = \frac{\partial L}{\partial \dot{\bf R}} = M\dot{\bf R},
\end{equation}    
is equal to the kinetic linear momentum.

The Hamiltonian function now follows from a Legendre transformation, giving
\begin{align}\label{eq:hamiltonian}
    H = &\; \frac{1}{2} \left ( {\bf J} + \frac{V_0}{\mu_0 \gamma_0}\chi {\bf B}_{\rm ext} \right )\cdot {\rm I}_{\rm eff}^{-1} \left ( {\bf J} + \frac{V_0}{\mu_0 \gamma_0}\chi {\bf B}_{\rm ext} \right ) \nonumber \\
    & +\frac{{\bf P}^2}{2M} - \frac{V_0}{2\mu_0} {\bf B}_{\rm ext} \cdot \chi {\bf B}_{\rm ext} - \frac{V_0}{12\mu_0} {\rm Tr}[(\xi : {\rm C}){\rm C}].
\end{align}
Note that here the angular momentum vector ${\bf J}$ must be interpreted as a function of the Euler angles and their canonically conjugate momenta $(p_\alpha,p_\beta,p_\gamma)$. For instance, for the body-frame components $J_k = {\bf J} \cdot {\bf n}_k$ one finds \cite{wachter2025gyroscopically},
\begin{subequations}\label{eq:canonicalangularmomentum2}
\begin{align}
J_1 =&\; p_\beta\sin\gamma-p_\alpha\csc\beta\cos\gamma+ p_\gamma\cot\beta \cos\gamma,\\
J_2 =&\; p_\beta\cos\gamma + p_\alpha\csc\beta\sin\gamma- p_\gamma\cot\beta \sin\gamma,\\
J_3 =&\; p_\gamma.
\end{align}
\end{subequations}

We will now use the Hamiltonian \eqref{eq:hamiltonian} as the starting point to (i) show how superconductors can fully align with the trapping field by calculating the normal-mode trapping frequencies, to (ii) discuss implications of continuous transfer of the angular momentum stored in the rigid-body rotation and in the supercurrents on the particle surface, and (iii) show how gyromagnetic coupling can be used to drive microparticle librations.

\begin{figure*}[t]
	\centering
	\includegraphics[width=\textwidth]{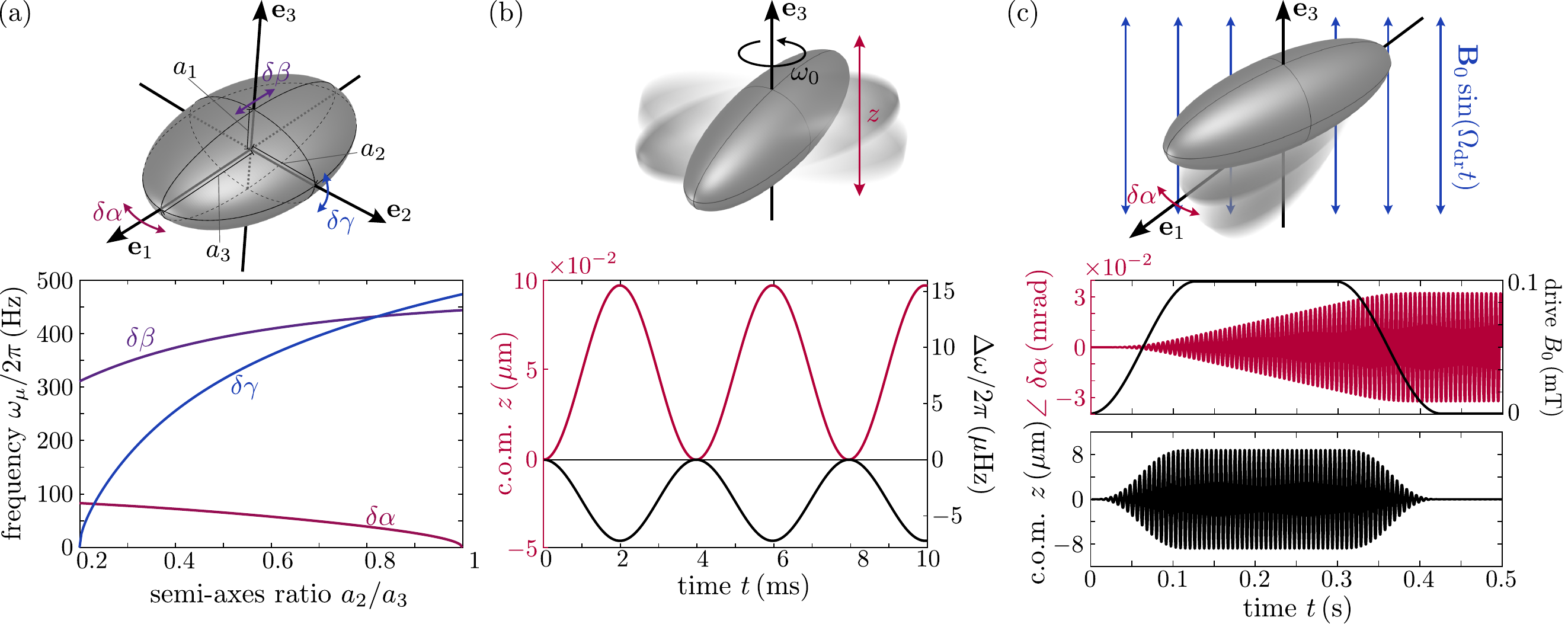}
	\caption{{\bf Full alignment \& gyromagnetic coupling:} (a) Librational frequencies obtained from the harmonic approximation of the trapping potential \eqref{eq:HarmonicExp} as a function of particle shape. The intermediate semi-axis is varied continuously from a prolate to an oblate spheroid, while the remaining semi-axes are held fixed. Analytic expression of the frequencies are given in Eq.~\eqref{eq:TrapFrequencies}. (b) Shift of the center-of-mass equilibrium position for a symmetric superconductor rotating about the ${\bf e}_3$ axis with angular velocity $\omega_0$. The particle is initially located at the equilibrium position corresponding to a non-rotating state. The shift arises from the Barnett-induced magnetization. (c) Gyromagnetic coupling between center-of-mass motion along ${\bf e}_3$ and the $\alpha$ libration. The translational motion is driven by an oscillating homogeneous magnetic field. The solid red curve shows the libration amplitude predicted by Eq.~\eqref{eq:drive_a} for $\omega_z / 2\pi = 161 \,{\rm Hz}$ and $\omega_\alpha / 2\pi = 173 \,{\rm Hz}$.}\label{fig:EdHDrive}
\end{figure*}

\subsection{Full alignment} \label{sec:3dalignment}

In a pure quadrupole field ${\bf B}_{\rm ext}({\bf r}) = {\rm C} {\bf r}$, the dipole potential in Eq.~\eqref{eq:lagrangian} describes stable trapping of the particle in the trap center ${\bf R} = 0$. If the particle shape is triaxial and for asymmetric trapping fields, the quadrupole potential in Eq.~\eqref{eq:hamiltonian} tends to align the principal axes with the external quadrupole field. Harmonically expanding the Hamiltonian around this stable equilibrium configuration yields the harmonic frequencies $\omega_{x,y,z}$ for small center-of-mass oscillations around the trap center and $\omega_{\alpha,\beta,\gamma}$ for small librations around the trap axes. The coupling between these normal modes by the Einstein-de Haas potential is negligible for realistic setups. The harmonically approximated Hamiltonian reads
\begin{align}\label{eq:HarmonicExp}
 H = &\; \sum_{i = 1}^3 \left ( \frac{\delta p^2_i}{2M}  + \frac{1}{2} M \omega_i^2 \delta x_i^2 \right ) \nonumber \\
 & + \sum_{\mu = \alpha,\beta,\gamma} \left ( \frac{\delta p_\mu^2}{2I_\mu} + \frac{1}{2} I_\mu \omega_\mu^2 \delta \mu^2 \right ).
\end{align}
Here, $\delta x_i$ and $\delta \mu$ denote small displacements from the equilibrium configuration with canonically conjugate momenta $\delta p_i$ and $\delta p_\mu$, respectively. Here, the equilibrium configuration is such that the longest particle axis aligns with the axis of smallest trapping gradient. Choosing the trapping gradients $|C_1| < |C_2| < |C_3|$ and the ellipsoid axes $a_1<a_2<a_3$ this means $(x_1,x_2,x_3) = (0,0,0)$ and $(\alpha,\beta,\gamma) = (0,\pi/2,0)$ . Thus, the ${\bf n}_1$ axis points along $-{\bf e}_z$, ${\bf n}_2$ along ${\bf e}_y$, and ${\bf n}_3$ along ${\bf e}_x$. Consequently, $\delta \alpha$ describes a small libration around the ${\bf e}_z$ axis with inertia $I_\alpha = I_1$, $\delta \beta$ describes librations around ${\bf e}_y$ with inertia $I_\beta = I_2$ and $\delta \gamma$ describes librations around ${\bf e}_x$ with inertia $I_\gamma = I_3$. Note that the equilibrium orientation $(\alpha, \beta,\gamma) = (0,\pi/2,0)$ is physically equivalent to the orientations obtained by rotating $\alpha$ or $\gamma$ by $\pi$ due to the inversion symmetry of the particle. 

Analytic expressions for the harmonic trapping frequencies can be found in Sec.~\ref{sec:frequencies}. They show that the trapping frequencies do not depend on the absolute size of the particle but only on its shape. Typical values of these frequencies for various aspect ratios are given in Tab.~\ref{tab:frequencies} and Fig.~\ref{fig:EdHDrive} (a), demonstrating that the center-of-mass frequencies are in the 100 Hz range. For markedly aspherical particles, the libration frequencies can clearly exceed the center-of-mass frequencies, reaching the $\sim 600$\,Hz range. This might well be an advantage for future precision experiments with levitated superconductors, where higher frequencies are advantageous. Sec.~\ref{sec:gravity} discusses the impact of Earth's gravity along the space-fixed $z$ direction on the equilibrium position and the trapping frequencies. These frequencies are relevant for ongoing experiments  with deeply trapped diamagnets \cite{hofer2022high,latorre2023}.

\begin{table}[b]
\caption{Trapping frequencies for differently shaped ellipsoidal superconductors in an external magnetic field with $C_1 = 58.1$\,T/m, $C_2 = 1.2 C_1$, and $C_3 = -2.2 C_1$, as motivated by Ref.~\cite{hofer2022high}. The frequencies are independent of the absolute size of the ellipsoid.}\label{tab:frequencies}
\begin{tabular}{ccccc}
    \hline\hline
	$a_1:a_2:a_3$ & $1:5:9$ & $1:2:3$ & $0.9:1:1.1$ & $1:1:3$\\\hline
	$\omega_x/2\pi$ (Hz) & $93$ & $98$ & $107$ & $95$\\
	$\omega_y/2\pi$  (Hz) & $117$ & $126$ & $132$ & $145$\\
	$\omega_z/2\pi$ (Hz) & $436$ & $304$ & $250$ & $266$\\
	$\omega_{\alpha}/2\pi$ (Hz) & $45$ & $40$ & $20$ & $58$\\
	$\omega_{\beta}/2\pi$ (Hz) & $518$ & $293$ & $97$ & $255$\\
	$\omega_{\gamma}/2\pi$ (Hz) & $466$ & $233$ & $64$ & $0$\\
    \hline \hline
\end{tabular}
\end{table}

\subsection{Gyromagnetic oscillations}

Let us now discuss the physical implications of the gyromagnetic coupling. As a first consequence, we will show that a spinning rotor can exhibit periodic oscillations in its center-of-mass motion due to the continuous interconversion of angular momentum between rigid-body rotation and magnetization. 

To see this, consider a symmetric particle $I_1 = I_2 \neq I_3$ in an azimuthally symmetric quadrupole trap with ${\rm C} = C ( {\bf e}_1\otimes {\bf e}_1 + {\bf e}_2\otimes{\bf e}_2 - 2 {\bf e}_3 \otimes {\bf e}_3)$. Due to these symmetries of the trap and the particle, the rotor dynamics preserve (i) the total energy \eqref{eq:hamiltonian}, (ii) the sum of orbital angular momentum and canonical angular momentum around the trap symmetry axis, and (iii) the angular momentum for rotations around the particle symmetry axis ${\bf n}_3$. 

The rotor is initially placed in the trap center $x=y=z=0$ with approximately $\dot{\gamma} = 0$ and $\beta = \pi/2$ and set in rotation around the trap-symmetry axis with frequency $\dot{\alpha} = \omega_0$. Without gyromagnetic coupling, the center of mass would remain at rest. However, due to the Barnett effect, the rotating particle exhibits an internal magnetization field, which leads to a force along the trap symmetry axis. This displaces the particle in $z$ direction and, as the particle moves away from the trap center, the trapping field reduces the induced dipole moment. The formation of the corresponding supercurrents slows the particle rotation due to the Einstein-de Haas effect. When the trapping field dominates over the Barnett field, the force is reversed and the particle moves back towards the trap center, see Fig. \ref{fig:EdHDrive} (b). This leads to a harmonic oscillation in the $z$ motion with amplitude 
\begin{equation}
    \Delta z  \approx \frac{\omega_0}{2 C \gamma_0}. 
\end{equation}
For typical trapping parameters and $\omega_0 \approx 1$\,MHz, this yields a displacement on the order of $\Delta z \approx 0.3\,\mu$m. This effect should be observable in state-of-the-art experiments \cite{latorre2023,hofer2022high}.

\subsection{Einstein-de Haas driving}\label{subsec:EdH_drive}

Gyromagnetic coupling can also be used to drive the librations with a periodic homogeneous field. Specifically, we now consider a symmetric particle in the asymmetric quadrupole trap  ${\rm C} = C \left[2 {\bf e}_1\otimes {\bf e}_1 - \left(1 - \delta/2\right) {\bf e}_2\otimes{\bf e}_2 - \left(1 + \delta/2\right) {\bf e}_3 \otimes {\bf e}_3\right]$. Initially, the particle is prepared in the potential minimum, so that it rests in the trap center $x = y = z = 0$ while its symmetry axis  ${\bf n}_3$ is aligned with the ${\bf e}_2$-axis so that $\alpha = \pi / 2$ and $\beta = \pi / 2$. In this configuration, there is no quadrupole torque on the particle.

If a periodic homogeneous field along the space-fixed $z$-direction is applied, this field induces a magnetic dipole moment along $z$ and thereby pushes the potential minimum out of the trap center. The particle then oscillates along $z$ around the new trap minimum. Along its path, it experiences no dipole-torque since the field and the dipole moment are parallel. However, as the magnetization changes, the particle starts librating in the $\alpha$ angle because of gyromagnetic coupling and the quadrupole torque. By periodically modulating the field in resonance with the libration frequency $\omega_\alpha$, the $\alpha$ rotation can be driven through gyromagnetic coupling. Specifically, for the driving field of amplitude $B_0$ and frequency $\Omega \approx \omega_\alpha$ applied for a duration $T \gg 2 \pi / |\omega_z - \Omega|$, where the process of switching on and off is slow compared to $\omega_z$ and short compared to $T$, we find for the maximal amplitude of the $\alpha$ libration,
\begin{align}\label{eq:drive_a}
    \alpha_{\rm max} \approx&\; \left| \frac{V_0 \chi_1 B_0 \omega_\alpha}{\mu_0 \gamma_0 I_1 \gamma_\alpha} \frac{\omega_\alpha - \omega_z}{(\omega_\alpha - \omega_z)^2 + \gamma_\alpha^2/4 } \right|.
\end{align}
Here, $\gamma_\alpha \ll |\omega_\alpha - \omega_z|$ is the mechanical damping of $\alpha$.

In Fig. \ref{fig:EdHDrive} (c), we show the amplitude of the driving field, the angle $\alpha$, and the center-of-mass position $z$ as a function of time for a prolate superconductor with semi-axes of $38\, {\rm \mu m}$ and $60\, {\rm \mu m}$ and a trap asymmetry of $\delta = 2/5$ with $C = 58.1 \,{\rm T m^{-1}}$. These parameters result in the frequencies of $\omega_z / 2\pi = 161 \,{\rm Hz}$ and $\omega_\alpha / 2\pi = 173 \,{\rm Hz}$. The amplitude of the driving field is $B_0 = 0.1 \,{\rm mT}$ and is applied for a duration of $T \approx 0.2\,{\rm s}$. While the background-field is switched on, the center-of-mass motion displays an oscillation with frequency $\omega_\alpha / 2\pi$ and reaches an amplitude of $8\,{\rm \mu m}$. The libration amplitude increases linearly to the maximal amplitude Eq.~\eqref{eq:drive_a}. This example illustrates that gyromagnetic coupling can be used to rotate levitated superconductors with available experimental techniques.

\subsection{Discussion}

In summary, we showed that induced quadrupolar diamagnetic torques and gyromagnetic coupling are crucial for understanding and controlling the rotational dynamics of aspherical diamagnetic objects in magnetic quadrupole traps. We showed that the quadrupolar contribution to the diamagnetic torque dominates close to the trap center, serving to three-dimensionally align the particle's main axes with the trap axes. Moreover, we showed how the gyromagnetic coupling of the particle rotation to the internally induced magnetization field can transfer angular momentum from the supercurrents, that shield the external magnetic fields from the particle's interior, to the particle rotation.

Our findings pave the way for controlling levitated superconductors in all six motional degrees of freedom, perhaps even in the quantum regime. This is a prerequisite for fully exhausting their technological and scientific potential for sensing and gravimetry \cite{pratcamps2017,timberlake2019,vinante2020,lewandowski2021,fuchs2024,Carney2025,headly2026quantum},  for probing hypothetical dark-matter models \cite{moore2021,Higgins2024}, and for quantum superposition tests \cite{bose2017,marletto2017,pino2018,millen2020b,lami2024testing,higgins2024superrot,bulling2026stability}. In addition, we proposed a scheme for the first experimental observation of gyromagnetic coupling in levitated superconducting particles. In addition to experiments with levitated superconductors, the devised strategies can be used to control the rotational motion of other magnetic and magnetizable objects.

\section{Rank-four demagnetization tensor}\label{sec:rank4demag}

To calculate the induced dipole vector, one uses Dirichlet's statement \cite{dittrich2016dirichlet} that the demagnetization tensor
\begin{align}\label{eq:N2relation}
{\rm N} = -\frac{1}{4\pi} \nabla_{\bf s} \otimes \nabla_{\bf s} \int_V d^3{\bf r}\, \frac{1}{|{\bf r} - {\bf s}|} = \sum_{i = 1}^3 N_i {\bf n}_i \otimes {\bf n}_i.
\end{align}
is independent of ${\bf s}$ and diagonal in the main-axes frame of the ellipsoid. The quantities $N_i$, known as demagnetization factors, can be explicitly given in terms of Jacobi elliptic functions \cite{stoner1945xcvii,osborn1945demagnetizing}.

To calculate the induced quadrupole tensor, we generalize the above statement by showing that the rank-three tensor
\begin{equation}\label{eq:N3def}
    {\rm D} = -\frac{1}{4\pi} \int_V d^3 {\bf r}\, ({\bf r} - {\bf R}) \otimes \nabla_{\bf s} \otimes \nabla_{\bf s} \frac{1}{|{\bf r} - {\bf s}|}
\end{equation}
is linear in ${\bf s} -{\bf R}$,
\begin{equation}\label{eq:N3relation}
    {\rm D} = \eta \cdot ({\bf s} - {\bf R}).
\end{equation}
Here the dot product between a rank-four tensor ${\cal A} = {\bf e}_i\otimes {\bf e}_j \otimes {\bf e}_k \otimes {\bf e}_\ell$ and a vector ${\bf b}$ is taken to contract in the right-most tensor component ${\cal A}\cdot {\bf b} = ({\bf b}\cdot {\bf e_\ell) \bf e}_i\otimes {\bf e}_j \otimes {\bf e}_k$. The proportionality is given by the rank-four tensor
\begin{equation}\label{eq:defeta}
    \eta = -\frac{1}{4\pi} \int_V d^3{\bf r}\, \frac{{\bf r}- {\bf R}}{|{\bf r} - {\bf s}|} \otimes \overleftarrow{\nabla}_{\bf s} \otimes \overleftarrow{\nabla}_{\bf s}\otimes \overleftarrow{\nabla}_{\bf s},
\end{equation}
which is independent of ${\bf s}$ and which we calculate analytically below. Here, $\overleftarrow{\nabla}_{\bf s}$ is the gradient with respect to ${\bf s}$ acting to the left. Importantly, we will see in Sec.~\ref{sec:derivationinducedmoments} that this defines the rank-four demagnetization tensor
\begin{equation} \label{eq:N4}
 {\cal N} =    \frac{1}{2} \left[\eta + \eta^{T_l} - \frac{2}{3} \mathds{1} \otimes {\rm Tr}_l \left(\eta\right)\right],
\end{equation}
where $T_l$ denotes the transposition of the two leftmost indices, $(\eta^{T_l})_{ijkl} = \eta_{jikl}$, and ${\rm Tr}_l(\cdot)$ is the partial trace over the two leftmost indices.

In order to demonstrate Eq.~\eqref{eq:N3relation} and calculate Eq.~\eqref{eq:defeta}, we will first reproduce Dirichlet's argument in modern language as presented in Ref.~\cite{dittrich2016dirichlet} and then generalize it to rank-four tensors.

\subsection{Rank-two demagnetization tensor}\label{sec:demag}

The rank-two demagnetization tensor can be evaluated by first transforming the integral to the reference position ${\bf R} = 0$ and in the reference orientation $\Omega = 0$ by substituting ${\bf r} = {\bf R} + {\rm R} {\bf x}$. We thus obtain
\begin{equation}
{\rm N} = -\frac{1}{4\pi} \nabla_{\bf s} \otimes \nabla_{\bf s} \int_{{\bf x}  \cdot {\rm S} {\bf x} \leq 1} d^3{\bf x}\, \frac{1}{|{\bf x} + {\rm R}^T ({\bf R} - {\bf s})|},
\end{equation}
where the shape tensor
\begin{equation}
{\rm S} = \sum_{j = 1}^3 \frac{1}{a_j^2}{\bf e}_j \otimes {\bf e}_j    
\end{equation}
determines the volume $V = \{{\bf r} \in \mathbb{R}^3|{\bf r} \cdot {\rm S} {\bf r} \leq 1\}$ of the ellipsoid of semi-axes lengths $a_j$ at the center-of-mass position ${\bf R} = 0$ and the reference orientation $\Omega = 0$, where the principal axes frame is aligned with the space frame. We proceed by multiplying the right-hand side with a Heaviside function
\begin{equation}\label{eq:dirichlet}
    \frac{1}{\pi} \int_C dz\, \frac{\sin z}{z} e^{i z \tau} = \begin{cases}
        1 & {\rm for} \quad |\tau| < 1 \\
        \frac{1}{2} &{\rm for}\quad  |\tau| = 1 \\
        0 & {\rm for} \quad |\tau| > 1.
    \end{cases}
\end{equation}
For later use, we here deform the integration domain to the  complex contour ${\cal C} = \{z = \zeta| \zeta \in \mathbb{R} \textbackslash [-\varepsilon,\varepsilon]\} \cup \{z = \varepsilon e^{-i \zeta}|\zeta\in [\pi,2\pi] \}$ with $\varepsilon > 0$; it lies along the real axis and excludes the origin at $z = 0$ by circling it via the upper-half of the complex plane.
Choosing for the parameter $\tau = {\bf x} \cdot {\rm S}{\bf x}$ yields
\begin{equation}\label{eq:demagderiv1}
    {\rm N} = -\frac{1}{4\pi^2} \nabla_{\bf s} \otimes \nabla_{\bf s} \int d^3{\bf x} \int_{\cal C} dz\, \frac{\sin z}{z} \frac{e^{i z {\bf x} \cdot {\rm S} {\bf x}}}{|{\bf x} + {\rm R}^T({\bf R} - {\bf s})|}.
\end{equation}
Note that the volume integral over ${\bf x}$ now covers all of $\mathbb{R}^3$.

This expression can be further evaluated by using the Euler formula
\begin{equation}
    \frac{\Gamma(y)}{x^y} e^{i \pi y/2} = \int_0^\infty d\kappa\, \kappa^{y-1} e^{i x \kappa},
\end{equation}
for $y = 1/2$, allowing us to write
\begin{align}\label{eq:euler}
    \frac{1}{|{\bf x} + {\rm R}^T({\bf R} - {\bf s})|} = &\; \frac{e^{-i \pi/4}}{\sqrt{\pi}} \int_0^\infty\, \frac{d\kappa }{\sqrt{\kappa}} \nonumber\\
    & \times \exp \left ( i \kappa |{\bf x} + {\rm R}^T({\bf R} - {\bf s})|^2 \right ).
\end{align}
Inserting this expression into Eq.~\eqref{eq:demagderiv1}, one can carry out the Gaussian integral over ${\bf x}$,
\begin{align}\label{eq:gaussianintegral}
    \int d^3{\bf x} &\, e^{ i {\bf x} \cdot (\kappa \mathds{1} + z{\rm S}) {\bf x}} e^{2 i \kappa {\rm R}^T ({\bf R} - {\bf s}) \cdot {\bf x}}  = e^{i3\pi/4}\sqrt{\frac{\pi^{3}}{{\rm det} ( \kappa \mathds{1} + z{\rm S})}} \nonumber \\
    & \times \exp \left [- i\kappa^2 ({\bf R} - {\bf s}) \cdot {\rm R}(\kappa \mathds{1} + z{\rm S})^{-1}{\rm R}^T  ({\bf R} - {\bf s})\right ]
\end{align}
yielding
\begin{align}
{\rm N} = &\;  -\frac{i}{4\pi}\nabla_{\bf s} \otimes \nabla_{\bf s} \int_C dz\, \frac{\sin z}{z} \int_0^\infty \frac{d\kappa}{\sqrt{\kappa}} {\rm det}( \kappa \mathds{1} + z {\rm S})^{-1/2} \nonumber \\
& \times \exp \left [i\kappa z ({\bf R}-{\bf s}) \cdot {\rm R} {\rm S} (\kappa \mathds{1}+ z {\rm S})^{-1}{\rm R}^T  ({\bf R}-{\bf s}) \right ].
\end{align}
Substituting $\kappa = z/\lambda$ gives
\begin{align}\label{eq:demagderiv2}
    {\rm N} = &\;  -\frac{i}{4\pi} \nabla_{\bf s} \otimes \nabla_{\bf s} \int_0^\infty \frac{d\lambda}{\sqrt{{\rm det}( \mathds{1} + \lambda {\rm S})}} \int_{\cal C} dz\, \frac{\sin z}{z^2}   \nonumber \\
& \times \exp \left [i z ({\bf R}-{\bf s}) \cdot {\rm R} \frac{{\rm S}}{\mathds{1}+ \lambda {\rm S}}{\rm R}^T ({\bf R}-{\bf s}) \right ].
\end{align}

The contour integration over $z$ can now be calculated explicitly by using the residue theorem. For this we note that $({\bf R} - {\bf s}) \cdot {\rm R}{\rm S} {\rm R}^T ({\bf R} - {\bf s}) \leq 1$, provided that ${\bf s}$ lies within the particle volume. In addition, since $\lambda \geq 0$, it also follows that $\delta = ({\bf R} - {\bf s}) \cdot {\rm R}{\rm S} (\mathds{1} + \lambda {\rm S})^{-1} {\rm R}^T ({\bf R} - {\bf s}) \leq 1$. The $z$ integral can be decomposed into two contributions
\begin{equation}
    \int_{\cal C} dz\, \frac{\sin z}{z^2} e^{iz \delta} = \frac{1}{2i} \int_{\cal C} \frac{dz}{z^2} e^{iz (1 + \delta)} - \frac{1}{2i} \int_{\cal C} \frac{dz}{z^2} e^{-iz (1 - \delta)}.
\end{equation}
For the first term, the contour can be closed in the upper complex plane by adding the path ${\cal C}_+ = \{z = \rho e^{i \xi}|\xi \in [0,\pi]\}$ with $\rho \to \infty$. Since the contour ${\cal C} + {\cal C}_+$ contains no singularity, the resulting integral vanishes identically. For the second term, the contour can be closed by adding the path ${\cal C}_- = \{z = \rho e^{-i \xi}|\xi \in [0,\pi]\}$ with $\rho \to \infty$. This path yields a finite contribution given by the residue theorem
\begin{equation}
    \int_{\cal C} dz\, \frac{\sin z}{z^2} e^{iz \delta} = -\frac{1}{2i} \int_{{\cal C} + {\cal C}_-} \frac{dz}{z^2} e^{-iz (1 - \delta)} = -i \pi (1 - \delta).
\end{equation}
Note that the negative sign is due to the negative orientation of the path ${\cal C} + {\cal C}_-$. Inserting this into Eq.~\eqref{eq:demagderiv2} yields
\begin{align}
    {\rm N} = &\; - \frac{1}{4} \nabla_{\bf s} \otimes \nabla_{\bf s} \int_0^\infty d\lambda\, {\rm det}( \mathds{1} + \lambda {\rm S})^{-1/2}\nonumber\\
    & \times \left [ 1 - ({\bf R}-{\bf s}) \cdot {\rm R}\frac{{\rm S}}{\mathds{1}+ \lambda {\rm S}} {\rm R}^T ({\bf R}-{\bf s})\right ].
\end{align}
Here, the derivatives can finally be evaluated, yielding
\begin{equation}\label{eq:demagN0}
    {\rm N} = \frac{1}{2}\int_{0}^\infty \frac{d\lambda}{\sqrt{{\rm det} (\mathds{1} + \lambda {\rm S})}}{\rm R} \frac{{\rm S}}{\mathds{1} + \lambda {\rm S}} {\rm R}^T.
\end{equation}
The demagnetization tensor is thus diagonal in the body frame,
\begin{equation}
{\rm N} = \sum_{i = 1}^3 N_i {\bf n}_i \otimes {\bf n}_i    
\end{equation}
with the three eigenvalues
\begin{align}\label{eq:demag}
    N_i = &\;  \frac{a_1a_2a_3}{2} \int_{0}^\infty\frac{d\lambda}{a_i^2 + \lambda} \prod_{k = 1}^3\frac{1}{\sqrt{a_k^2 + \lambda}}.
\end{align}
They are known as the demagnetization coefficients.

We can calculate the trace over the demagnetization tensor \eqref{eq:demagN0} by using Jacobi's formula
\begin{equation}
    \frac{d}{d \lambda} {\rm det} ({\rm A}) = {\rm det}({\rm A}) {\rm Tr} \left ({\rm A}^{-1} \frac{d}{d \lambda}{\rm A} \right ),
\end{equation}
which holds for any invertible tensor ${\rm A}$. Choosing ${\rm A} = \mathds{1} + \lambda {\rm S}$ yields
\begin{equation}
    {\rm Tr}({\rm N}) = -\int_0^\infty d\lambda\, \frac{d}{d\lambda} \frac{1}{\sqrt{{\rm det}(\mathds{1} + \lambda {\rm S})}} = 1,
\end{equation}
thus showing that the sum of the demagnetization factors, i.e. the eigenvalues of Eq.~\eqref{eq:demagN0} fulfill
\begin{equation}
    N_1 + N_2 + N_3 = 1.
\end{equation}
This implies that the tensor is fully characterized by only two body-fixed elements.

\subsection{Rank-four demagnetization tensor}\label{sec:demag4}

We now have everything required to derive the rank-four demagnetization tensor $\eta$. As a first step, we substitute ${\bf r} = {\bf R} + {\rm R} {\bf x}$ in Eq.~\eqref{eq:N3def} and multiply the right-hand side with Eq.~\eqref{eq:dirichlet} to obtain
\begin{align}\label{eq:N3deriv1}
    {\rm D} = &\; -\frac{1}{4\pi^2} \int_{{\bf x}\cdot {\rm S} {\bf x}} d^3  {\bf x} \int_{\cal C} dz\, \frac{\sin z}{z} \frac{({\rm R}{\bf x}) e^{i z {\bf x}\cdot {\rm S} {\bf x}}}{|{\bf x} + {\rm R}^T ({\bf R} - {\bf s})|}\nonumber \\
    & \otimes \overleftarrow{\nabla}_{\bf s} \otimes \overleftarrow{\nabla}_{\bf s}.
\end{align}
Here, $\overleftarrow{\nabla}_{\bf s}$ indicates that the derivative operators act to the left. The volume integral over ${\bf x}$ can again be performed by applying Euler's formula \eqref{eq:euler}, so that we are confronted with the Gaussian integral
\begin{align}
    \int & d^3{\bf x}\, {\rm R}{\bf x} e^{ i {\bf x} \cdot (\kappa \mathds{1} + z{\rm S}) {\bf x}+2 i \kappa {\rm R}^T ({\bf R} - {\bf s}) \cdot {\bf x}}  =\pi^{3/2} \frac{ie^{i3\pi/4}}{2\kappa}\\
    & \times \nabla_{\bf s}\frac{\exp \left [- i\kappa^2 ({\bf R} - {\bf s}) \cdot {\rm R}(\kappa \mathds{1} + z{\rm S})^{-1}{\rm R}^T  ({\bf R} - {\bf s})\right ]}{\sqrt{{\rm det} ( \kappa \mathds{1} + z{\rm S})}},\nonumber
\end{align}
where we replaced ${\rm R} {\bf x}$ by $ i \nabla_{\bf s}/2\kappa$ under the integral and used Eq.~\eqref{eq:gaussianintegral}. Carrying out the derivative and inserting the result in Eq.~\eqref{eq:N3deriv1} yields
\begin{align}\label{eq:F4}
    {\rm D} = &\;  {\bf f}({\bf s})\otimes \overleftarrow{\nabla}_{\bf s} \otimes \overleftarrow{\nabla}_{\bf s}
\end{align}
with
\begin{align}
    {\bf f}({\bf s} )= &\; \frac{i}{4\pi} \int_0^\infty d\kappa\, \sqrt{\kappa} \int_{\cal C} dz\, \frac{\sin z}{z} {\rm det} (\kappa \mathds{1} + z{\rm S})^{-1/2}\nonumber \\
    & \times \exp \left [i\kappa^2 ({\bf R} - {\bf s}) \cdot {\rm R}(\kappa \mathds{1} + z{\rm S})^{-1} {\rm S} {\rm R}^T  ({\bf R} - {\bf s})\right ]\nonumber \\
    & \times {\rm R} (\kappa \mathds{1} + z {\rm S})^{-1} {\rm R}^T ({\bf R}- {\bf s}).
\end{align}
Substituting $\lambda = z/\kappa$ yields
\begin{align}
    {\bf f}({\bf s}) = &\; \frac{i}{4\pi} \int_0^\infty  \frac{d\lambda}{\sqrt{{\rm det}(\mathds{1} + \lambda {\rm S})}} \int_{\cal C} dz\, \frac{\sin z}{z^2} \nonumber \\
& \times \exp \left [i z ({\bf R}-{\bf s}) \cdot {\rm R} \frac{{\rm S}}{\mathds{1}+ \lambda {\rm S}}{\rm R}^T ({\bf R}-{\bf s}) \right ]\nonumber \\
& \times {\rm R} (\mathds{1} + \lambda {\rm S})^{-1} {\rm R}^T ({\bf R}- {\bf s}).
\end{align}
Again, the integral along ${\cal C}$ can be calculated with the residue theorem by using the same paths as above. This yields
\begin{align}
    {\bf f}({\bf s}) = &\; \frac{1}{4} \int_0^\infty  \frac{d\lambda}{\sqrt{{\rm det}(\mathds{1} + \lambda {\rm S})}} \nonumber \\
     & \times \left [1 - ({\bf R} - {\bf s}) \cdot {\rm R} \frac{{\rm S}}{\mathds{1} + \lambda {\rm S}} {\rm R}^T ({\bf R} - {\bf s}) \right ]\nonumber\\
     & \times {\rm R} (\mathds{1} + \lambda {\rm S})^{-1} {\rm R}^T ({\bf R}- {\bf s}).
\end{align}
Calculating the derivatives in \eqref{eq:F4} shows that indeed
\begin{equation}
{\rm D} =  \eta\cdot({\bf s} - {\bf R})
\end{equation}
with
\begin{align}
    \eta = &\; \eta_{ij} \left[{\bf n}_i \otimes {\bf n}_j \otimes \left({\bf n}_j \otimes {\bf n}_i + {\bf n}_i \otimes {\bf n}_j \right)\right.\nonumber\\
    &\left. + {\bf n}_i \otimes {\bf n}_i \otimes {\bf n}_j \otimes {\bf n}_j\right],
\end{align}
which is totally symmetric in its three rightmost indices and ${\rm Tr}_r \left(\eta\right) = \mathds{1}$ (with ${\rm Tr}_r$ the partial trace over the two rightmost indices). The elements of $\eta$ are
\begin{equation}\label{eq:demag4}
 {\eta}_{ij} = \frac{a_1a_2a_3}{2} \int_0^\infty \frac{d\lambda \, a_i^2}{(a_i^2 + \lambda)(a_j^2 + \lambda)} \prod_{k = 1}^3\frac{1}{\sqrt{a_k^2 + \lambda}}.
\end{equation}
A straight-forward calculation shows that they fulfill the transposition rule
\begin{equation}
    \eta_{ij}  = \eta_{ji} + N_j - N_i
\end{equation}
and the sum rule
\begin{equation}
 2 \eta_{ii} + \sum_{j = 1}^3 \eta_{ij} = 1,
\end{equation} 
which also implies
\begin{equation}
    2 \eta_{jj} + \sum_{i = 1}^3 \eta_{ij} = 3 N_j,
\end{equation}
for $i = 1, 2, 3$.

The rank-four demagnetization tensor Eq.~\eqref{eq:N4} respects several symmetries: the quadrupole tensor and the magnetic field gradient tensor are symmetric tensors, demanding that ${\cal N}_{ijkl} = {\cal N}_{jikl} = {\cal N}_{ijlk}$. These two relations reduce the number of independent elements from 81 to 36. Magnetostatic reciprocity requires ${\cal N}_{ijkl} = {\cal N}_{klij}$, further reducing the number of independent elements to 21. Moreover, triaxial ellipsoids possess three reflection symmetries so that all elements ${\cal N}_{ijkl}$ where each axis appears an odd number of times must vanish. This implies that the tensor can have at most nine independent elements. Finally, accounting for the fact that ${\cal N}$ must be traceless with respect to the two leftmost and the two rightmost indices shows that there are only six independent entries.

In order to account for this symmetry, we define the coefficients
\begin{align}
    N_{ij}^{(1)} = &\; \frac{\eta_{ij}+\eta_{ji}}{2} \nonumber \\
    = &\; \frac{a_1a_2a_3}{4} \int_0^\infty \frac{d\lambda \, (a_i^2 + a_j^2)}{(a_i^2 + \lambda)(a_j^2 + \lambda)} \prod_{k = 1}^3\frac{1}{\sqrt{a_k^2 + \lambda}}.
\end{align}
and
\begin{align}
    N_{ij}^{(2)} = &\; N_j - \eta_{ij}\nonumber \\
    = &\; \frac{a_1 a_2 a_3}{2} \int_0^\infty \frac{d\lambda \, \lambda}{(a_i^2+\lambda)(a_j^2+\lambda)} \prod_{k=1}^3 \frac{1}{\sqrt{a_k^2+\lambda}}.
\end{align}
which are symmetric $N_{ij}^{(k)} = N_{ji}^{(k)}$ and fulfill the relation
\begin{equation}
    2N^{(1)}_{ii} - \sum_{j=1}^3 N^{(2)}_{ij} = 0.
\end{equation}
They allow us to compactly write the rank-four demagnatization tensor as
\begin{align}\label{eq:demag4tensor}
    {\cal N} = &\; \sum_{i,j = 1}^3 N_{ij}^{(1)} {\bf n}_i \otimes {\bf n}_j \otimes \left( {\bf n}_i \otimes {\bf n}_j + {\bf n}_j \otimes {\bf n}_i\right) \nonumber\\
    & - \sum_{i,j = 1} ^3 N_{ij}^{(2)} {\bf n}_i \otimes {\bf n}_i \otimes {\bf n}_j \otimes {\bf n}_j.
\end{align}
A straightforward calculation shows that the components ${\cal N}_{ijk\ell} =
({\bf n}_i \otimes {\bf n}_j) : {\cal N}(\Omega) : ({\bf n}_k \otimes {\bf n}_\ell)$ are symmetric, ${\cal N}_{ijk\ell} = {\cal N}_{jik\ell} = {\cal N}_{ij\ell k} = {\cal N}_{\ell kji}$. In addition, the rank-four demagnatization tensor is left-traceless ${\rm Tr}_l({\cal N}) = 0$ and right-traceless ${\rm Tr}_r({\cal N}) = 0$. Note that this implies that the rank-four demagnetization tensor can indeed be written through only six independent body-fixed elements
\begin{align}\label{eq:demag4final}
    {\cal N} = &\; \sum_{i,j = 1 \atop {i\neq j}}^3 N_{ij}^{(1)} {\bf n}_i \otimes {\bf n}_j \otimes \left( {\bf n}_i \otimes {\bf n}_j + {\bf n}_j \otimes {\bf n}_i\right) \nonumber\\
    & + \sum_{i,j = 1\atop {i\neq j}} ^3 N_{ij}^{(2)} {\bf n}_i \otimes {\bf n}_i \otimes \left({\bf n}_i \otimes {\bf n}_i-{\bf n}_j \otimes {\bf n}_j \right).
\end{align}

\section{Induced quadrupole tensor}\label{sec:derivationinducedmoments}

Having extended Dirichlet's relation \eqref{eq:N2relation} to Eq.~\eqref{eq:N3relation}, we are now in the position to determine the induced quadrupole tensor inside an ellipsoidal particle. In a first step, we will derive the integral equation \eqref{eq:intmag} for the magnetization field, then in a second step use it to calculate the induced dipole vector \eqref{eq:magdip}, and finally derive the induced quadrupole tensor \eqref{eq:magquad}.

\subsection{Magnetization field integral equation}\label{sec:inducedmagn}

Let us now turn to the integral equation \eqref{eq:intmag} for the magnetization field ${\bf M}({\bf r})$ induced by the external field ${\bf B}_{\rm ext}({\bf r})$ inside a magnetizable object of permeability $\mu_r({\bf r})$. In the absence of free charges, the total magnetic field ${\bf B}({\bf r})$ follows Maxwell's equations of magnetostatics, 
\begin{align}
\nabla\cdot{\bf B}({\bf r})&=0&\nabla\times{\bf B}({\bf r})&=\mu_{0}[{\bf j}_f({\bf r}) + {\bf j}_{b}({\bf r})]
\end{align}
where the bound charge current is ${\bf j}_{b}=\nabla\times{\bf M}$, and the free current gives rise to the external field,
\begin{equation}
    {\bf B}_{\rm ext}({\bf r}) = \frac{\mu_0}{4\pi} \nabla \times \int d^3{\bf s}\, \frac{{\bf j}_f({\bf s})}{|{\bf r} - {\bf s}|}.
\end{equation}
These equations can be solved by ${\bf B} = \nabla \times {\bf A}$ with the vector potential
\begin{equation}
    {\bf A}({\bf r}) = \frac{\mu_0}{4\pi} \int d^3{\bf s}\, \frac{1}{|{\bf r} - {\bf s}|} \left [ {\bf j}_f({\bf s}) + \nabla_{\bf s} \times {\bf M}({\bf s}) \right ].
\end{equation}
Using Gauss theorem on the second term on the right-hand side yields the magnetic field
\begin{equation}
    {\bf B}({\bf r}) = {\bf B}_{\rm ext}({\bf r}) + \frac{\mu_0}{4\pi} \left [ \nabla \otimes \nabla - \Delta \mathds{1} \right ]  \int d^3{\bf s}\, \frac{{\bf M}({\bf s})}{|{\bf r} - {\bf s}|}.
\end{equation}
This expression can be rewritten by using $\Delta 1/|{\bf r} - {\bf s}| = -4\pi \delta({\bf r}-{\bf s})$ and by identifying the static Green tensor ${\rm G}({\bf r}) = \nabla \otimes \nabla 1/4\pi |{\bf r}|$. We can eliminate the magnetic field from this equation with the help of the constitutive relation ${\bf M}({\bf r}) = [\mu_r({\bf r})-1] [{\bf B}({\bf r}) - \boldsymbol{\omega}/\gamma_0]/\mu_0\mu_r({\bf r})$, which follows from \eqref{eq:constitutiverotating} for single species magnetism with gyromagnetic ratio $-\gamma_0$. This yields the integral equation Eq.~\eqref{eq:intmag}.

\subsection{Induced dipole vector} 

In the next step, we use the integral equation \eqref{eq:intmag} for the quadrupole field \eqref{eq:quadfield} to derive expressions for the induced dipole vector. Specifically, using Eq.~\eqref{eq:magdipdef} and integrating Eq.~\eqref{eq:magnetization} with  ${\bf B}_0({\bf s}) = {\bf B}_0({\bf R}) + {\rm C}({\bf s} - {\bf R})$ we can write the induced dipole vector as
\begin{equation}
    {\bf m} =\frac{1}{\mu_0} \int_V d^3{\bf r} \int_V d^3{\bf s}\, {\rm K}({\bf r}, {\bf s}) \left [ {\bf B}_0({\bf R}) + {\rm C} ({\bf s} - {\bf R}) \right ].
\end{equation}
Defining the rank-two susceptibility tensor
\begin{equation}\label{eq:defchi2}
    \chi = \frac{1}{V_0} \int_V d^3{\bf r} \int_V d^3{\bf s}\, {\rm K}({\bf r},{\bf s}),
\end{equation}
and the rank-three susceptibility tensor
\begin{equation}\label{eq:defchi3}
    \xi_3 = \frac{1}{V_0} \int_V d^3 {\bf r} \int_V d^3{\bf s}\, {\rm K}({\bf r},{\bf s}) \otimes ({\bf s} - {\bf R}),
\end{equation}
the dipole vector can thus be written as
\begin{equation}
    {\bf m} = \frac{V_0}{\mu_0} \left [ \chi {\bf B}_0({\bf R}) + \xi_3 : {\rm C}\right ].
\end{equation}

The two tensors $\chi$ and $\xi_3$ can be calculated through the integral equation \eqref{eq:inteqk}. Before doing so, note that the symmetries discussed in Sec.~\ref{sec:demag4} already imply that $\xi_3$ must vanish for ellipsoidal bodies. Specifically, the three reflection symmetries of triaxial ellipsoids imply that the rank-two susceptibility is diagonal in the body-fixed frame while the rank-three tensor vanishes. This also follows from a direct calculation by inserting the definition \eqref{eq:defchi2} into Eq.~\eqref{eq:inteqk}, yielding
\begin{equation}\label{eq:chi2eq}
    \chi = (\mu_r - 1) \left [\mathds{1} - \frac{1}{V_0} \int_V d^3{\bf r} d^3{\bf s} d^3{\bf s}' \, {\rm G}({\bf r} - {\bf s'}) {\rm K}({\bf s}',{\bf s}) \right ].
\end{equation}
We can now carry out the ${\bf r}$-integration and identify the demagnetization tensor \eqref{eq:N2relation} from Dirichlet’s relation. The remaining integrals yield the susceptibility tensor \eqref{eq:defchi2}. Since both tensors are diagonal in the same basis we can solve for Eq.~\eqref{eq:chi2eq}, yielding Eq.~\eqref{eq:chi2} with
\begin{equation}\label{eq:chii}
    \chi_i = \frac{\mu_r - 1}{1 + (\mu_r - 1) N_i}.
\end{equation}
Likewise, we can calculate the rank-three tensor \eqref{eq:defchi3} by using Eq.~\eqref{eq:inteqk}, so that
\begin{equation}
\xi_3 = \frac{\mu_r - 1}{V_0} \int_V d^3 {\bf r}d^3 {\bf s} d^3 {\bf s}'\, {\rm G}({\bf r} - {\bf s}') {\rm K}({\bf s},{\bf s}') \otimes ({\bf s} - {\bf R}).
\end{equation}
Here, the first term in Eq.~\eqref{eq:inteqk} does not contribute due to the point-symmetry of the ellipsoid. Using the demagnetization tensor yields a closed equation for $\xi_3$, whose solution is indeed $\xi_3 = 0$.

\subsection{Induced quadrupole tensor}

The above calculation can be generalized to determine the quadrupole tensor \eqref{eq:Qdef}. Inserting the field \eqref{eq:quadfield} together with relation \eqref{eq:magnetization} yields
\begin{equation}\label{eq:Qij}
    {\rm Q} = \frac{3 V_0}{\mu_0} \left[\sigma + \sigma^T - \frac{2}{3} \mathds{1} \otimes {\rm Tr}_l \left(\sigma\right) \right] : {\rm C}.
\end{equation}
Here we used that $\xi_3 = 0$ and we define the rank-four tensor
\begin{equation}
    \sigma = \frac{1}{V_0} \int_V d^3{\bf r} \int_V d^3{\bf s} \, ({\bf r} - {\bf R}) \otimes {\rm K}({\bf r},{\bf s}) \otimes ({\bf s} - {\bf R}).
\end{equation}
Its body-frame components fulfill $\sigma_{ijkl} = \sigma_{lkji}$ because of the symmetry of the integral kernel \eqref{eq:inteqk}. Inserting Eq.~\eqref{eq:inteqk} gives
\begin{align}\label{eq:eqxi4}
    \sigma = &\; \frac{\mu_r - 1}{V_0} \left[\int_V d^3{\bf r}\, ({\bf r} - {\bf R}) \otimes \mathds{1} \otimes ({\bf r} - {\bf R})\nonumber\right.\\
    & + \int_V d^3{\bf r} d^3{\bf s} d^3{\bf s'}\, ({\bf r} - {\bf R})\nonumber\\
    &\left.\vphantom{\int_V} \otimes \left[{\rm G}({\bf r} - {\bf s}') {\rm K}({\bf s}',{\bf s})\right] \otimes ({\bf s} - {\bf R}) \right].
\end{align}
We can now use the result \eqref{eq:N3relation} for ellipsoidal bodies to rewrite this as a linear tensor equation
\begin{equation}\label{eq:sig}
    \sigma = \left(\mu_r - 1\right) \left({\cal M} - \eta : \sigma\right),
\end{equation}
where
\begin{align}
    {\cal M} = &\; \frac{1}{V_0} \int_V d^3{\bf r}\, ({\bf r} - {\bf R}) \otimes \mathds{1} \otimes ({\bf r} - {\bf R}).
\end{align}

Equation \eqref{eq:Qij} determines how the rank-four susceptibility \eqref{eq:xi4} is given in terms of $\sigma$, 
\begin{align} \label{eq:xisig}
    \xi = &\; \frac{3}{2} \left[\sigma + \sigma^{T_l} + \sigma^{T_r} + \left(\sigma^{T_l}\right)^{T_r}\right] - \mathds{1} \otimes {\rm Tr}_l\left(\sigma +\sigma^{T_r}\right)\nonumber\\
    &- {\rm Tr}_r \left(\sigma + \sigma^{T_l}\right) \otimes \mathds{1} + \frac{2}{3}{\rm Tr}\left[{\rm Tr}_l \left(\sigma\right)\right] \mathds{1} \otimes \mathds{1}.
\end{align}
Here, the last two terms ensure that the rank-four susceptibility tensor obeys the symmetries  $\xi_{ijkl} = \xi_{jikl} = \xi_{ijlk} = \xi_{klij}$ and that it is traceless with respect to the two right-most indices. Note that the two terms can be added without modifying the quadrupole tensor since the field gradient tensor ${\rm C}$ is symmetric and traceless. These symmetries imply that the rank-four susceptibility tensor must be of the form
\begin{align}
    \xi\left(\Omega\right) = &\; \sum_{i,j = 1 \atop i \ne j}^{3} \left[\xi_{ij}^{(1)} {\bf n}_i \otimes {\bf n}_j \otimes \left({\bf n}_i \otimes {\bf n}_j + {\bf n}_j \otimes {\bf n}_i\right)\right.\nonumber \\
    &\left.\hspace{1cm} + \xi_{ij}^{(2)} {\bf n}_i \otimes {\bf n}_i \otimes \left( {\bf n}_i \otimes {\bf n}_i - {\bf n}_j \otimes {\bf n}_j\right)\right].
\end{align}
In the following we will derive expressions for the six parameters $\xi_{ij}^{(1)}$ and $\xi_{ij}^{(2)}$.

Inserting Eq.~\eqref{eq:sig} into Eq.~\eqref{eq:xisig} gives the linear equation for the rank-four susceptibility tensor
\begin{equation}\label{eq:xiequation}
    \xi = \left(\mu_r - 1\right) \left({\cal W} - {\cal N} : \xi\right).
\end{equation}
Here, we used the definition of the rank-four demagetization tensor \eqref{eq:N4} and we defined the symmetric second mass moment tensor
\begin{align}\label{eq:mtensor}
    {\cal W} = &\; \frac{3}{2} \left[{\cal M} + {\cal M}^{T_l} + {\cal M}^{T_r} + \left({\cal M}^{T_l} \right)^{T_r}\right]\nonumber\\
    &-\mathds{1} \otimes {\rm Tr}_l \left( {\cal M} + {\cal M}^{T_r}\right)- {\rm Tr}_r \left({\cal M} + {\cal M}^{T_l}\right) \otimes \mathds{1} \nonumber\\
    & + \frac{2}{3} {\rm Tr} \left[{\rm Tr}_r \left({\cal M}\right)\right] \mathds{1} \otimes \mathds{1}.
\end{align}
It can be calculated by direct integration and takes the form
\begin{align}
    {\cal W} = &\; \sum_{i,j=1 \atop i\ne j}^3  \left[w_{ij}^{(1)} {\bf n}_i \otimes {\bf n}_j \otimes \left({\bf n}_i \otimes {\bf n}_j + {\bf n}_j \otimes {\bf n}_i\right)\right.\nonumber \\
    &\left.\hspace{1cm} + w_{ij}^{(2)} {\bf n}_i \otimes {\bf n}_i \otimes \left( {\bf n}_i \otimes {\bf n}_i - {\bf n}_j \otimes {\bf n}_j\right)\right],
\end{align}
with
\begin{align}
    w_{ij}^{(1)} = &\; \frac{3 (a_i^2 + a_j^2 )}{10}\\
    w_{ij}^{(2)} = &\; \frac{2}{15} \left[3 (a_i^2 + a_j^2) - \sum_{k=1}^3 a_k^2\right].
\end{align}

Equation \eqref{eq:xiequation} can be solved by collecting the terms that contract with $\xi$ and invert them with respect to the double-dot product,
\begin{align}\label{eq:xi4formal}
    \xi\left(\Omega\right) = &\; \left(\mu_r - 1\right) \left[ \mathds{1}_4 + \left(\mu_{r} - 1\right) {\cal N}(\Omega) \right]^{-1} : {\cal W}(\Omega).
\end{align}
$\mathds{1}_4$ denotes the rank-four identity tensor
\begin{equation}
	\mathds{1}_{4}=\sum_{i,j=1}^{3}{\bf n}_{i}\otimes {\bf n}_{j}\otimes{\bf n}_{i}\otimes{\bf n}_{j},
\end{equation}
so that $\mathds{1}_4 : {\cal A} = {\cal A}$ for arbitrary ${\cal A}$. 

The inverse tensor
\begin{equation}
    {\cal K} = (\mu_r - 1) \left [ \mathds{1}_4 + (\mu_r - 1) {\cal N}\right ]^{-1}
\end{equation}
can be calculated by noting the block structure of ${\cal N}$, see Eq.~\eqref{eq:demag4final}, as
\begin{align}
    {\cal K} = &\; (\mu_r - 1) \sum_{i=1}^3 {\bf n}_i \otimes {\bf n}_i\otimes {\bf n}_i \otimes {\bf n}_i \nonumber\\
    & + (\mu_r-1)\sum_{i,j=1\atop i\neq j}^3 {\bf n}_i \otimes {\bf n}_j \otimes \left  ({\bf n}_i \otimes {\bf n}_j -{\bf n}_j\otimes {\bf n}_i \right )\nonumber\\
    & + \sum_{i,j=1\atop i\neq j}^3 K^{(1)}_{ij} {\bf n}_i \otimes {\bf n}_j \otimes \left  ({\bf n}_i \otimes {\bf n}_j +{\bf n}_j\otimes {\bf n}_i \right )\nonumber \\
    & - \sum_{i,j=1\atop i\neq j}^3 K_{ij}^{(2)} {\bf n}_i \otimes {\bf n}_i \otimes \left( {\bf n}_i \otimes {\bf n}_i - {\bf n}_j \otimes {\bf n}_j \right),
\end{align}
where
\begin{equation}
    K_{ij}^{(1)} = \frac{\mu_r - 1}{\left[1 + 2 \left(\mu_r - 1\right) N_{ij}^{(1)}\right]},
\end{equation}
and
\begin{equation}
    K_{ij}^{(2)} = \frac{(\mu_r - 1)^2}{{\rm det} (K)} \left[N_{ij}^{(2)} + (\mu_r - 1) \nu'\right],
\end{equation}
with
\begin{align}
    \nu = &\; N_{12}^{(2)} + N_{23}^{(2)} + N_{31}^{(2)},\\
    \nu' = &\; N_{12}^{(2)}N_{13}^{(2)} + N_{12}^{(2)}N_{23}^{(2)} + N_{23}^{(2)}N_{13}^{(2)},\\
    {\rm det} (K) = &\; 1 + 2 (\mu_r-1) \nu + 3(\mu_r-1)^2 \nu'.
\end{align}

This implies for the coefficients of the rank-four susceptibility tensor Eq.~\eqref{eq:xi4}
\begin{align}
    \xi_{ij}^{(1)} = &\; \frac{3}{10} \frac{\left(\mu_r - 1\right) \left(a_i^2 + a_j^2\right)}{\left[1 + 2 \left(\mu_r - 1\right) N_{ij}^{(1)}\right]}
\end{align}
as well as
\begin{align}
    \xi_{ij}^{(2)} = &\; \frac{2 (\mu_r - 1)}{15} \left[ 1 - \frac{3 (\mu_r - 1)^2 \nu'}{{\rm det} (K)} \right] \left[ 2(a_i^2 + a_j^2) - a_k^2 \right] \nonumber\\
    & - \frac{(\mu_r - 1)^2}{5~ {\rm det} (K)} \left[ 3(a_i^2 + a_j^2) N_{ij}^{(2)} + (a_i^2 - a_k^2) N_{ik}^{(2)}\right.\nonumber\\
    &\left.+ (a_j^2 - a_k^2) N_{jk}^{(2)} \right],
\end{align}
where $(ijk)$ are even permutations of (123).

\section{Dynamics of trapped superconductors}\label{sec:dynamics}

We can now identify the Lagrangian that generates the Newtonian equations of motion  \eqref{eq:eom}. This Lagrangian is then the starting point for a Legendre transformation to discuss the phase-space dynamics of trapped superconductors.

\subsection{Lagrangian equations of motion} \label{sec:lagrangian}

In order to identify the Lagrangian from the equation of motions \eqref{eq:eomfinal}, we convert them into Euler-Lagrange form. For the center of mass position, a straightforward calculation shows that the equation of motion \eqref{eq:eom} with induced dipole vector \eqref{eq:magdip} implies
\begin{align}\label{eq:comlag}
    \left(\frac{d}{d t} \frac{\partial}{\partial \dot{\bf R}} - \frac{\partial}{\partial {\bf R}}\right) \left[ \frac{M}{2} \dot{\bf R}^2 + \frac{V_0}{\mu_0 \gamma_0} \boldsymbol{\omega} \cdot \chi {\bf B}_{\rm ext}\left({\bf R}\right)\right.&\nonumber\\
    \left. + \frac{V_0}{2 \mu_0} {\bf B}_{\rm ext}\left({\bf R}\right) \cdot \chi {\bf B}_{\rm ext}\left({\bf R}\right) \right] &= 0.
\end{align}
For the rotational motion, we first need to express the equations of motion in terms of a concrete parametrization $\Omega = (\alpha,\beta,\gamma)$. For this we choose Euler angles in the $z$-$y'$-$z''$ convention. The rotation tensor ${\rm R}(\Omega)$ then rotates the space-frame axes to the body-fixed unit vectors ${\bf n}_k(\Omega) = {\rm R}(\Omega) {\bf e}_k$ given by
\begin{subequations}
\begin{equation}
    {\bf n}_1 = \left ( \begin{array}{c}
         \cos\alpha \cos \beta \cos \gamma - \sin \alpha \sin \gamma\\
         \sin \alpha \cos \beta \cos \gamma + \cos \alpha \sin \gamma \\
         -\sin \beta \cos \gamma
    \end{array}\right ),
\end{equation}
\begin{equation}
    {\bf n}_2 = \left ( \begin{array}{c}
         -\cos\alpha \cos \beta \sin \gamma - \sin \alpha \cos \gamma\\
         -\sin \alpha \cos \beta \sin \gamma + \cos \alpha \cos \gamma \\
         \sin \beta \sin \gamma
    \end{array}\right ),
\end{equation}
\begin{equation}
    {\bf n}_3 = \left ( \begin{array}{c}
         \cos\alpha \sin \beta\\
         \sin \alpha \sin \beta \\
         \cos \beta
    \end{array}\right ).
\end{equation}
\end{subequations}
Here, vector elements refer to the space-fixed basis ${\bf e}_k$. The angular velocity vector takes the form $\boldsymbol{\omega} = \dot{\alpha}{\bf e}_z + \dot{\beta} {\bf e}_n + \dot{\gamma} {\bf n}_3$, with the nodal line ${\bf e}_n = -\sin\alpha {\bf e}_x + \cos \alpha {\bf e}_y$. Its body-frame components are
\begin{subequations} \label{eq:omegaanglederivatives}
\begin{align}
    \omega_1 & = {\bf n}_1 \cdot \boldsymbol{\omega} = \dot{\beta} \sin \gamma - \dot{\alpha} \sin \beta \cos \gamma, \\
    \omega_2 & = {\bf n}_2 \cdot \boldsymbol{\omega} = \dot{\beta} \cos \gamma + \dot{\alpha} \sin \beta \sin \gamma, \\
    \omega_3 & = {\bf n}_3 \cdot \boldsymbol{\omega} = \dot{\alpha} \cos \beta + \dot{\gamma}.
\end{align}    
\end{subequations}
For the following calculation, it is helpful to note the relations for $\mu = \alpha, \beta, \gamma$ \cite{martinetz2022surface}
\begin{subequations}\label{eq:luk}
\begin{align}
\label{eq:luk1}
\frac{\partial}{\partial \dot{\mu}} = &\; {\bf m}_\mu \cdot \frac{\partial}{\partial \boldsymbol{\omega}}, \\
\label{eq:luk2}
\frac{\partial {\rm R}}{\partial \mu} = &\; {\bf m}_\mu \times {\rm R}, \\
\label{eq:luk3}
\frac{d {\bf m}_\mu}{dt} = &\; {\rm R} \frac{\partial}{\partial \mu} {\rm R}^T \boldsymbol{\omega}.
\end{align}
\end{subequations}
Here, we abbreviated ${\bf m}_\alpha = {\bf e}_3$, ${\bf m}_\beta = {\bf e}_{\rm n}$, and ${\bf m}_\gamma = {\bf n}_3$.

We start by multiplying the left-hand side of Eq.~\eqref{eq:eomfinal2} with ${\bf m}_\mu$ and using Eqs.~\eqref{eq:luk} this becomes
\begin{align}\label{eq:Lag1}
    & \left(\frac{d}{d t} \frac{\partial}{\partial \dot{\mu}} - \frac{\partial}{\partial \mu} \right) \left( \frac{1}{2} \boldsymbol{\omega} \cdot {\rm I}_{\rm eff} \boldsymbol{\omega} - \frac{V_0}{2 \mu_0 \gamma_0} \boldsymbol{\omega} \cdot \chi {\bf B}_{\rm ext}\right)\nonumber\\
    & + \frac{V_0}{2 \mu_0 \gamma_0} \frac{\partial \boldsymbol{\omega}}{\partial \mu} \cdot \chi {\bf B}_{\rm ext}.
\end{align}

The scalar product on the right-hand side of Eq.~\eqref{eq:eomfinal2} with ${\bf m}_\mu$ reads
\begin{equation}
    \frac{V_0}{\mu_0} \left[{\bf m}_\mu \times \chi \left({\bf B}_{\rm ext} - \frac{\boldsymbol{\omega}}{\gamma_0} \right)\right] \cdot {\bf B}_{\rm ext} + \frac{V_0}{3 \mu_0}\left({\bf m}_\mu \times \xi : {\rm C} \right) : {\rm C}.
\end{equation}
This can be rewritten by using that the angle derivatives of the susceptibility tensors follow from using \eqref{eq:luk2} for each basis vector,
\begin{equation}
    \frac{V_0}{2 \mu_0} \frac{\partial \chi}{\partial \mu} \left( {\bf B}_{\rm ext} - \frac{\boldsymbol{\omega}}{\gamma_0}\right) \cdot {\bf B}_{\rm ext} + \frac{V_0}{12 \mu_0} \left(\frac{\partial \xi}{\partial \mu} : {\rm C} \right) : {\rm C}.
\end{equation}
Together with Eq.~\eqref{eq:Lag1} this allows us to write the equations of motion as
\begin{align}
    \left(\frac{d}{d t} \frac{\partial}{\partial \dot{\mu}} - \frac{\partial}{\partial \mu} \right) \left[ \frac{1}{2} \boldsymbol{\omega} \cdot {\rm I}_{\rm eff} \boldsymbol{\omega} + \frac{V_0}{\mu_0 \gamma_0} \chi \boldsymbol{\omega} \cdot {\bf B}_{\rm ext}\right.&\nonumber\\
    \left. + \frac{V_0}{2 \mu_0} {\bf B}_{\rm ext} \cdot \chi {\bf B}_{\rm ext} + \frac{V_0}{12 \mu_0} \left(\xi : {\rm C} \right) : {\rm C} \right] &= 0.
\end{align}
Together with Eq.~\eqref{eq:comlag}, we thus find that the equations of motion \eqref{eq:eomfinal} are generated by the Lagrangian \eqref{eq:lagrangian}.

To derive the Hamiltonian, we have to calculate the canonical momenta. While the  center-of-mass canonical momentum is given by the linear kinetic momentum \eqref{eq:canonicallinearmomentum}, the canonical angular momentum contains a gyromagnetic contribution,
\begin{equation}
    p_\mu = \frac{\partial L}{\partial \dot{\mu}} = {\bf m}_\mu \cdot \left ( {\rm I}_{\rm eff} \boldsymbol{\omega}_0 -  \frac{V_0}{\mu_0\gamma_0} \chi {\bf B}_{\rm ext}\right ).
\end{equation}
Here, we used relation \eqref{eq:luk1} and, as above, the right-hand-side is understood in terms of the Euler angles and their time derivatives \eqref{eq:omegaanglederivatives}. We can now identify the canonical momentum vector through $p_\mu = {\bf J} \cdot {\bf m}_\mu$, implying relations \eqref{eq:canonicalangularmomentum} as well as \eqref{eq:canonicalangularmomentum2}. A Legendre transformation yields the Hamiltonian \eqref{eq:hamiltonian}.

\subsection{Harmonic trapping frequencies}\label{sec:frequencies}

In order to obtain the harmonic frequencies of small-amplitude center-of-mass oscillations and angular librations, we first explicitly state the potentials in the Hamiltonian \eqref{eq:hamiltonian} for the quadrupole field ${\bf B}_{\rm ext}({\bf R}) = {\rm C} {\bf R}$. Denoting $x_i = {\bf R} \cdot {\bf e}_i$, the dipole potential 
\begin{equation}
    V_{\rm dip} = -\frac{V_0}{2 \mu_0} {\bf B}_{\rm ext} \cdot \chi {\bf B}_{\rm ext}
\end{equation}
is
\begin{equation}
    V_{\rm dip} = -\frac{V_0}{2\mu_0} \sum_{i,j,k = 1}^3 \chi_k C_i C_j ({\bf n}_k \cdot {\bf e}_i)({\bf n}_k \cdot {\bf e}_j)x_i x_j.
\end{equation}
Here, we used that the quadrupole tensor of the trapping field is diagonal in the space-fixed frame ${\rm C} {\bf e}_i = C_i {\bf e}_i$ with eigenvalues $C_i$. Likewise, the quadrupole potential 
\begin{equation}
     V_{\rm quad} =  -\frac{V_0}{12 \mu_0} \left(\xi : {\rm C} \right) : {\rm C}
\end{equation}
can be written as
\begin{align}\label{eq:quadpotexp}
    V_{\rm quad} = &\; - \frac{V_0}{12 \mu_0} \sum_{i,j,k,\ell = 1}^3C_i C_j (\xi_{k\ell}^{(1)} + \xi_{\ell k}^{(2)})  \nonumber\\
    & \times ({\bf n}_k \cdot {\bf e}_i) ({\bf n}_\ell \cdot {\bf e}_i) ({\bf n}_k \cdot {\bf e}_j) ({\bf n}_\ell \cdot {\bf e}_j) \nonumber \\
    & - \frac{V_0}{12 \mu_0} \sum_{i,j,k,\ell = 1}^3 C_i C_j \xi_{k \ell}^{(2)} ({\bf n}_k \cdot {\bf e}_i)^2 ({\bf n}_\ell \cdot {\bf e}_j)^2.
\end{align}

Considering an asymmetric ellipsoid with $a_1 < a_2 < a_3$ in a trapping field characterized by the quadrupole parameters $|C_1| < |C_2| < |C_3|$, the harmonic trapping frequencies are obtained by Taylor expanding the sum of these potentials around the equilibrium configuration $x_i = 0$ and $\Omega = (0,\pi/2,0)$. This yields
\begin{subequations}\label{eq:TrapFrequencies}
\begin{align}
    \omega^2_x = &\; \frac{V_0 |\chi_3| C_1^2}{M \mu_0}, \\
    \omega^2_y = &\; \frac{V_0 |\chi_2| C_2^2}{M \mu_0}, \\
    \omega^2_z = &\; \frac{V_0 |\chi_1| C_3^2}{M \mu_0}, \\
    \omega_{\alpha}^2 =&\; \frac{V_0 \left|C_1 - C_2\right|}{3 \mu_0 I_1} \left| 2 \left(C_1 - C_2\right) \left(\xi_{23}^{(2)} - \xi_{23}^{(1)}\right)\right.\nonumber\\
    &\left.- \left( C_2 - C_3 \right) \xi_{12}^{(2)} - \left( C_3 - C_1 \right) \xi_{13}^{(2)} \right|, \\
    \omega_{\beta}^2 =&\; \frac{V_0 \left|C_3 - C_1\right|}{3 \mu_0 I_2} \left| 2 \left(C_3 - C_1\right) \left(\xi_{13}^{(2)} - \xi_{13}^{(1)}\right)\right.\nonumber\\
    &\left.- \left( C_1 - C_2 \right) \xi_{23}^{(2)} - \left( C_2 - C_3 \right) \xi_{12}^{(2)} \right|, \\
    \omega_{\gamma}^2 =&\; \frac{V_0 \left|C_2 - C_3\right|}{3 \mu_0 I_3} \left| 2 \left(C_2 - C_3\right) \left(\xi_{12}^{(2)} - \xi_{12}^{(1)}\right)\right.\nonumber\\
    &\left. - \left( C_3 - C_1 \right) \xi_{13}^{(2)} - \left( C_1 - C_2 \right) \xi_{23}^{(2)}\right| .
\end{align}
\end{subequations}

\subsection{Gyromagnetic oscillations}\label{sec:symmrotor}

We now discuss the impact of the gyromagnetic coupling on the particle motion. For this we consider a symmetric particle with $I_1 = I_2 = I$ in an azimuthally symmetric quadrupole trap with $C_1 = C_2 = C$ and $C_3 = - 2C$. Choosing cylindrical coordinates $(\rho,\varphi,z)$ for the center of mass ${\bf R} = \rho (\cos \varphi {\bf e}_x + \sin\varphi {\bf e}_y) + z {\bf e}_z$, the kinetic energy of the Lagrangian \eqref{eq:lagrangian} is
\begin{align}
    T_{\rm kin} = &\; \frac{M}{2} \left(\dot{\rho}^2 + \rho^2 \dot{\varphi}^2 + \dot{z}^2\right) + \frac{I}{2} \left(\dot{\alpha}^2 \sin^2 \beta + \dot{\beta}^2\right) \nonumber \\
     & + \frac{I_3}{2} (\dot{\alpha}\cos\beta + \dot{\gamma})^2.
\end{align}    
The symmetry of the particle also implies that $\chi_1 = \chi_2 = \chi_\bot$ and $\chi_3 = \chi_\|$ so that the generalized potential of gyromagnetic coupling
\begin{equation}
    V_{\rm gyr} = \frac{V_0}{\mu_0 \gamma_0} \boldsymbol{\omega} \cdot \chi {\bf B}_{\rm ext} =  \frac{V_0}{\mu_0 \gamma_0} \sum_{i,j = 1}^3 \chi_j C_i ({\bf n}_j \cdot {\bf e}_i) \omega_j x_i,
\end{equation}
becomes
\begin{align}\label{eq:eidhcoupling}
    V_{\rm gyr} = &\; \frac{V_0 C \chi_\bot}{\mu_0\gamma_0} \left [ \rho \dot{\beta} \sin (\varphi - \alpha) + \rho \dot{\gamma} \sin \beta \cos (\varphi - \alpha) \right.\nonumber \\
    & \left. - 2 z (\dot{\alpha} + \cos \beta \dot{\gamma} ) \vphantom{\dot{\beta} \sin (\varphi - \alpha)} \right ] \nonumber \\
    & + \frac{V_0 C \Delta \chi}{\mu_0\gamma_0} (\dot{\alpha}\cos \beta  + \dot{\gamma}) \left[ \rho \sin \beta \cos(\varphi - \alpha)\right.\nonumber\\
    & \left. - 2 z \cos \beta\right ]
\end{align}
where $\Delta \chi = \chi_\| - \chi_\bot$. Likewise, the dipole potential is
\begin{align}
    V_{\rm dip} = &\; - \frac{V_0 \chi_\bot C^2}{2 \mu_0} ( \rho^2 + 4z^2) \nonumber\\
     & - \frac{V_0 \Delta \chi C^2}{2 \mu_0} [ \rho \sin \beta \cos( \varphi - \alpha) -2z \cos\beta]^2.
\end{align}
In order to determine the quadrupole potential we use that for symmetric rotors, 
\begin{align}
\xi_{13}^{(i)} = \xi_{23}^{(i)}  \textrm{ for $i = 1,2$} .
\end{align}
Using these relations in Eq.~\eqref{eq:quadpotexp} yields
\begin{align}\label{eq:vquad}
    V_{\rm quad} = &\; \frac{3 V_0 C^2}{8 \mu_0} \left( \xi_{12}^{(2)} - \xi_{13}^{(2)} \right) \cos 2 \beta \nonumber\\
    & - \frac{3 V_0 C^2}{32 \mu_0} \left(\xi_{12}^{(2)} + 5 \xi_{13}^{(2)} - 4 \xi_{13}^{(1)}\right) \cos 4 \beta,
\end{align}
up to a constant energy shift. The first term yields a harmonic trapping potential close to $\beta = \pi/2$ where $\cos^2 \beta \approx (\beta - \pi/2)^2$. The symmetry axis of the rotor is thus approximately confined to the symmetry plane of the quadrupole trap.

Note that $\gamma$ is cyclic, implying that the angular momentum
\begin{align}
    p_\gamma = &\; I_3 (\dot{\alpha}\cos \beta + \dot{\gamma}) \nonumber\\
    & + \frac{V_0 C \chi_\|}{\mu_0\gamma_0} [ \rho \sin \beta \cos(\varphi - \alpha) - 2 z \cos \beta]
\end{align}
is conserved. In addition, the Lagrange function depends only on $\varphi - \alpha$ and is thus invariant under global rotations around the space-fixed $z$-axis. The associated angular momentum
\begin{align}
    L_z =&\; M\rho^2 \dot{\varphi} + I \sin^2 \beta \dot{\alpha} + p_\varphi \cos \beta \nonumber\\
    & - \frac{V_0 C \Delta\chi}{\mu_0\gamma_0} \cos \beta  \left[ \rho \sin \beta \cos(\varphi - \alpha)- 2 z \cos \beta\right ]\nonumber \\
    & -  \frac{2V_0 C \chi_\bot}{\mu_0\gamma_0} z
\end{align}
is thus also conserved.

We now consider the situation that the particle is initially located at the trap center $\rho_0 = 0$ and $z_0 = 0$, oriented to lie within the trapping plane $\beta_0 = \pi/2$, not rotating around its symmetry axis $\dot{\gamma}_0 = 0$ but spinning around the trap axis $\dot{\alpha}_0 = \omega_0$, so that $p_\varphi = 0$ and $L_z = I \omega$. Then, the transverse particle position will remain at the trap center $\rho = 0$, the particle symmetry axis will remain orthogonal to the trap symmetry axis $\beta = \pi/2$, and there will be no eigenrotation $\dot{\gamma} =0$ for all times. The dynamics thus reduces to the $\alpha$-rotation around the trap symmetry axis and the $z$-translation along the trap symmetry axis
\begin{subequations}
\begin{align}
    M \ddot{z} =&\; - \frac{2 V_0 \chi_\bot C }{\mu_0 \gamma_0} \dot{\alpha} + \frac{4V_0\chi_\bot C^2}{\mu_0}z\\
    I \ddot{\alpha} = &\;  \frac{2 V_0 \chi_\bot C }{\mu_0 \gamma_0} \dot{z},
\end{align}    
\end{subequations}
where the second equation expresses the conservation of angular momentum $L_z$. It can be used to eliminate $\dot{\alpha}$ from the first equation, i.e. using $\dot{\alpha} = \omega_0 + 2 V_0 \chi_\bot C z/\mu_0 \gamma_0$, so that we obtain
\begin{equation}
    M\ddot{z} \approx - \frac{2 V_0 \chi_\bot C }{\mu_0 \gamma_0}\omega_0 +  \frac{4V_0\chi_\bot C^2}{\mu_0} z,
\end{equation}
where we neglected terms of order $1/\gamma_0^2$. The gyromagnetic coupling due to the Einstein-de Haas effect thus shifts the equilibrium position of the $z$-motion by
\begin{equation}
    \Delta z \approx \frac{\omega_0}{2 C \gamma_0}. 
\end{equation}

\begin{figure*}[t]
	\centering
	\includegraphics[width=\textwidth]{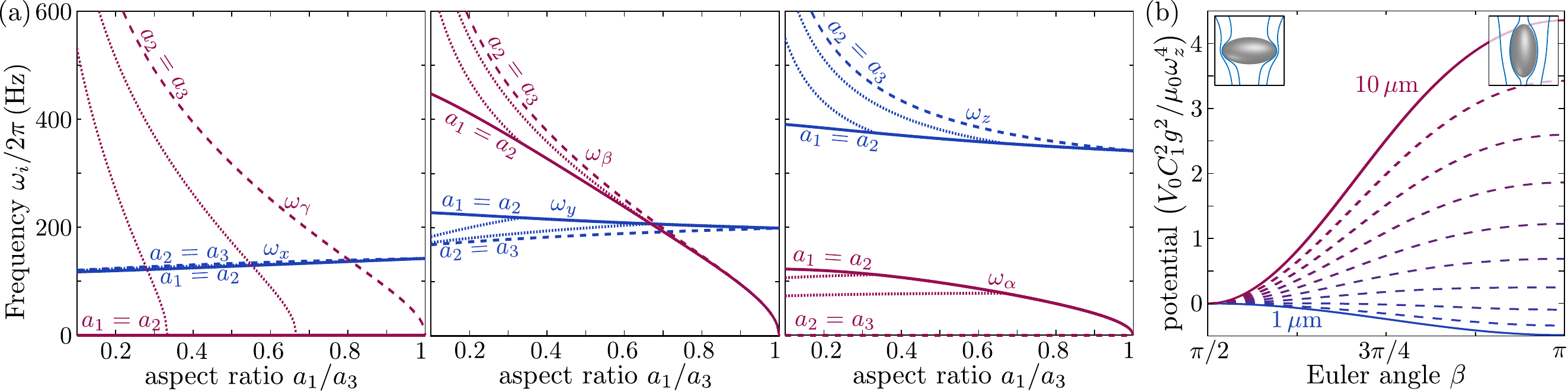}
	\caption{{\bf Trapping frequencies:} (a) Center-of-mass and librational frequencies as a function of the particle aspect ratio $a_1 / a_3$, where $a_1 \le a_2 \le a_3$. From left to right, the panels show the frequencies associated with motion along and libration about the ${\bf e}_1$, ${\bf e}_2$, and ${\bf e}_3$ axes. Solid curves correspond to prolate spheroids ($a_1 = a_2$), while dashed curves correspond to oblate spheroids ($a_2 = a_3$). Dotted curves indicate the transition between the two geometries for fixed ratios $a_2/a_3 = 1/3$ and $a_2/a_3 = 2/3$. All panels share the same frequency scale. (b) Orientation dependence of the trapping potential in presence of homogeneous gravity [second line of Eq.~\eqref{eq:potentialgravity}] as a function of Euler-angle $\beta$ for a prolate spheroid with increasing size. The insets show how the equilibrium alignment of the  semi-axis $a_3$ rotates by 90$^\circ$.}\label{fig:freq}
\end{figure*}

\subsection{Einstein-de Haas driving} \label{sec:spinrot}

We consider a symmetric particle $I_1 = I_2$ in an asymmetric quadrupole trap with 
\begin{align}
    {\rm C} = &\; C \bigg[2 {\bf e}_1\otimes {\bf e}_1 - \left(1 - \frac{\delta}{2}\right) {\bf e}_2\otimes{\bf e}_2\nonumber\\
    & - \left(1 + \frac{\delta}{2}\right) {\bf e}_3 \otimes {\bf e}_3\bigg].
\end{align}

The oscillating homogeneous magnetic driving field ${\bf B}_{\rm dr}(t) = \sin (\Omega_{\rm dr} t) B_{0} (t) {\bf e}_3$ of frequency $\Omega_{\rm dr}$ and with envelope $B_{0}(t)$ (of maximal amplitude $B_0$) induces a time-dependent magnetic dipole moment in the levitated particle. The quadrupole field leads to a force acting on the dipole, shifting the particle's equilibrium position along the magnetic field axis. At the same time, the librational mode around the magnetic field axis also starts librating upon changing the magnetic dipole moment due gyromagnetic coupling. The equations of motion for oscillations along and librations around the magnetic field axis are
\begin{subequations}
\begin{align}
    \ddot{z} = &\; - \omega_z^2 z - \gamma_z \dot{z} - \frac{V_0 \chi_1 C_3}{\mu_0 \gamma_0 m} \dot{\alpha}\nonumber\\
    & + \frac{V_0 \chi_1 C_3}{\mu_0 m} B_0(t) \sin (\Omega_{\rm dr} t ),\\
    \ddot{\alpha} &= - \omega_\alpha^2 \alpha - \gamma_\alpha \dot{\alpha} + \frac{V_0 \chi_1 C_3}{\mu_0 \gamma_0 I_1} \dot{z}\nonumber\\
    & - \frac{V_0 \chi_1}{\mu_0 \gamma_0 I_1} \frac{d}{d t} \left [ B_0(t) \sin (\Omega_{\rm dr} t) \right ].
\end{align}
\end{subequations}
For small libration amplitudes $\alpha$, we can neglect the impact of the librations on the center-of-mass motion. If the amplitude of the driving field varies slowly in comparison to the difference of center-of-mass and driving frequency, $\left|\Omega_z - \Omega_{\rm dr}\right|/2\pi \gg \dot{B}_0/B_0$, where $\Omega_z = \sqrt{\omega_z^2 - \gamma_z^2/4}$, the center-of-mass amplitude adiabatically follows the drive amplitude,
\begin{align}
    z(t) \approx &\; - \frac{V_0 C_3 \chi_1 }{2 \mu_0 m \Omega_{z}} \frac{B_0(t)}{(\Omega_{\rm dr} - \Omega_z)^2 + \gamma_z^2/4}\nonumber\\
    &\times \bigg[ (\Omega_{\rm dr} - \Omega_z) \sin (\Omega_{\rm dr} t) + \frac{\gamma_z}{2} \cos (\Omega_{\rm dr} t) \bigg].
\end{align}
Assuming that the driving field is nearly resonant with the libration mode, $\Omega_{\rm dr} \approx \Omega_\alpha = \sqrt{\omega_\alpha^2 - \gamma_\alpha^2 / 4 }$, we obtain
\begin{align}\label{eq:alphalong}
    \alpha(t) \approx &\; \frac{V_0 \chi_1}{2 \mu_0 \gamma_0 I_1}\nonumber\\
    & \times\Bigg[\bigg( 1 + \frac{\omega_z^2 ( \Omega_\alpha - \Omega_z - \gamma_\alpha \gamma_z / 4 \Omega_\alpha )}{\Omega_z \left[ (\Omega_\alpha - \Omega_z)^2 + \gamma_z^2 / 4 \right]} \bigg)\sin (\Omega_{\rm dr} t)\nonumber\\
    & + \frac{\gamma_\alpha}{2 \Omega_\alpha} \bigg(1 + \frac{\omega_z^2 ( \Omega_\alpha - \Omega_z - \Omega_\alpha \gamma_z / \gamma_\alpha )}{\Omega_z \left[ (\Omega_\alpha - \Omega_z)^2 + \gamma_z^2 / 4 \right]}\bigg) \cos (\Omega_{\rm dr} t)\Bigg]\nonumber\\
    & \int_{0}^{t} dt'\, B_0(t') e^{- \gamma_\alpha (t - t') / 2}.
\end{align}
If the drive amplitude $B_0(t) = B_0$ is constant for the duration $T$ and damping is small $\omega_{z,\alpha} \gg \gamma_{z,\alpha}$, Eq.~\eqref{eq:alphalong} can be further approximated, yielding the amplitude \eqref{eq:drive_a}.

\subsection{Impact of gravity}\label{sec:gravity}

Earth's gravitational field exerts a force that slightly pushes the particle out of the trap center. In order to estimate how gravity affects the particle trapping frequencies and normal modes, we add the gravitational potential $M g {\bf e}_3 \cdot {\bf R}$ to the Hamiltonian \eqref{eq:hamiltonian}, assuming that gravity points in the space-fixed direction $-{\bf e}_3$. The total potential can thus be written as
\begin{align}\label{eq:potentialgravity}
    V = &\; - \frac{V_0}{2 \mu_{0}} \left({\bf R} - {\bf R}_0\right) \cdot {\rm C} \chi {\rm C} \left({\bf R} -{\bf R}_0\right) \notag\\
		& - \frac{V_0}{2 \mu_{0}} \left[ {\rm Tr} \left [ \left(\xi : {\rm C}\right) {\rm C} \right ] - {\bf R}_0 \cdot \left({\rm C} \chi {\rm C}\right) {\bf R}_{0}\right],
\end{align}
with the shifted center-of-mass equilibrium position
\begin{equation}\label{eq:positiongravity}
	{\bf R}_{0} = \mu_0 g\frac{M}{V_0 }({\rm C} \chi {\rm C})^{-1} {\bf e}_3.
\end{equation}
The equilibrium position is proportional to the mass density of the particle and thus independent of the particle volume and size. The center-of-mass trapping frequencies, the corresponding normal modes, and the equilibrium position \eqref{eq:positiongravity} depend on the equilibrium particle orientation $\Omega_0$ through the tensor $\chi$. 

The equilibrium orientation $\Omega_0$ is determined by the minimization of the second line in Eq.~\eqref{eq:potentialgravity}. While the second term in parentheses does not depend on the absolute particle size, the first term scales as $V_0^{2/3}$ through the rank-four susceptibility tensor Eq.~\eqref{eq:xi4}. This gives rise to two limiting scenarios: (i) First, for small particles the second term dominates and the particle will align such that the equilibrium position \eqref{eq:positiongravity} becomes maximally negative. This implies that the minimal absolute susceptibility axis is aligned with ${\bf e}_3$ while the other two principal axes align with field gradients in the transverse direction. (ii) Second, for big particles the first term in the second line of Eq.~\eqref{eq:potentialgravity} dominates and the particle tends to align with the field gradient tensor as discussed in Sec.~\ref{sec:3dalignment}. The center-of-mass trapping frequencies are then as discussed in Eqs.~\eqref{eq:TrapFrequencies} and the gravity-induced shift in the center-of-mass position is determined by the susceptibility component in direction of gravity. In both limiting cases the equilibrium position is ${\bf R}_0 = - g {\bf e}_3/\omega_z^2$, where the value of $\omega_z$ is determined by which susceptibility axes is aligned with gravity. In summary, and perhaps counterintuitively, the overall impact of gravity is less for bigger objects, which tend to align with the field gradients. For instance, for prolate spheroids we find that the shift in the rotational frequencies only approaches single digit percentages for particle sizes below $\sim 60$\,$\mu$m.

The continuous transition between these two limiting scenarios is depicted in Fig.~\ref{fig:freq} (b) for a prolate particle in an azimuthally symmetric quadrupole field. The second line of the potential \eqref{eq:potentialgravity} then depends only on the polar angle $\beta$, see Eq.~\eqref{eq:vquad}.
\begin{equation}
    V_{\rm rot} = V_{\rm quad} + \frac{\mu_0 M^2 g^2}{8 \chi_\bot \chi_\| C^2} \left( \chi_\| - \Delta \chi\cos^2 \beta \right).
\end{equation}

\section{Summary and outlook}\label{sec:outlook}

The core theoretical innovations of this work are: (i) The analytic calculation of the induced magnetic moments as required to understand the diamagnetic response of ellipsoidal superconductors in inhomogneneous magnetic fields. (ii) The derivation of the equations of motion involving the diamagnetic quadrupole torque and gyromagnetic coupling, laying the foundation for controlling levitated superconducting rotors. (iii) Demonstrating how  superconducting ellipsoids can be fully aligned in asymmetric quadrupole traps. (iv) The identification of a viable strategy for the first experimental observation of strong gyromagnetic coupling in levitated superconductors. Our findings thus pave the way for controlling levitated superconductors in all six motional degrees of freedom, perhaps even in the quantum regime. This is a prerequisite for fully exhausting their technological and scientific potential for sensing and gravimetry \cite{pratcamps2017,timberlake2019,vinante2020,lewandowski2021,fuchs2024,Carney2025,headly2026quantum},  for probing hypothetical dark-matter models \cite{moore2021,Higgins2024}, and for quantum superposition tests \cite{bose2017,marletto2017,pino2018,millen2020b,lami2024testing,higgins2024superrot,bulling2026stability}.

In addition, our findings can be expected to have implications well beyond their immediate applications for levitated superconductors. We deem the following three aspects especially interesting:
\begin{enumerate}
    \item[(a)] The rotational dynamics of permanent magnetic particles with non-homogeneous magnetization fields, as discussed in Sec.~\ref{sec:magrot}, is relevant for ongoing experiments with Meissner-levitated ferromagnets above superconductors \cite{wang2019,gieseler2020singlespin,ahrens2026observation,wang2026exploring}. Specifically, in these particles a slight asymmetry can give rise to a finite magnetic quadrupole moment and thereby break the rotational symmetry around the magnetic dipole axis. This can lead to strong intermodal couplings as well as to three-dimensional alignment, as also observed experimentally \cite{wang2026exploring}.
    \item[(b)] The generalization of Dirichlet's method in Sec.~\ref{sec:rank4demag} and the calculation of the induced magnetic quadrupole tensor in Sec.~\ref{sec:derivationinducedmoments} can also be used to determine the electric quadrupole moment induced in an ellipsoidal dielectric nanoparticle that is illuminated by the field of an optical tweezer. Specifically, the polarization field induced in such a particle \cite{rudolph2021} resembles the magnetization field derived in Sec.~\ref{sec:derivationinducedmoments}, so that the methods presented here can be directly applied. A treatment beyond the Rayleigh approximation of dipole scattering is particularly relevant for ongoing experiments with micron-sized objects.
    \item[(c)] Gyromagnetic coupling from induced magnetic moments can also play a prominent role in other magnetizable systems, such as levitated diamonds \cite{perdriat2021spin}, naphtalene crystals \cite{steiner2025optically}, and paramagnetic particles \cite{pinilla2026unveiling}. The theoretical techniques developed here might well be applied to study the impact of gyromagnetic couplings on the particle dynamics in such systems.
\end{enumerate}

We expect our findings to serve as the starting point for several follow-up studies that generalize the presented theoretical treatment. For instance, a natural generalization of this work will be to extend it to account analytically and numerically for higher-order induced moments beyond the quadrupole. Such a description becomes empirically relevant if either the particle shape cannot be approximated as ellipsoidal or if the local trapping field is not well characterized by a homogeneous field gradient \cite{cunill2026macroscopic}. The resulting induced magnetization field will then be characterized by higher-order susceptibility tensors which describe how higher-order gradients of the trapping field induce higher-order moments of the magnetization field. Furthermore, experiments with smaller particles will require a quantitative treatment of the impact of the finite London penetration depth since such particles are not well approximated as perfect diamagnets. Extending the theory into this direction will require solving an integral equation for the magnetization field that explicitly accounts for the finite magnetic field in a thin layer close to the particle surface \cite{hofer2019analytic}. It may also be pertinent to generalize our description to type-II superconductors, whose frozen-in fluxes can strongly impact the coherent and incoherent particle dynamics \cite{brandt1988friction}. Quantifying the corresponding forces and torques, as well as the noise associated with thermally actived vortex jumps will be crucial for future quantum applications with such particles.  

\begin{acknowledgements}
    We thank Markus Aspelmeyer for discussions on observing gyromagnetic coupling with superconductors. F.K., B.A.S., and J.A. acknowledge funding by the Carl-Zeiss-Foundation through the Carl-Zeiss-Foundation Center QPhoton. B.A.S. is supported by the DFG No. 510794108 and 583862189 (QuantERA CoMaQLev). J.A. acknowledges financial support from the BMFTR through the Cluster4Future QSens (project QMat) and the German Science Foundation (DFG) through AN336/18-1. W.W. acknowledges support by the Horizon Europe 2021-2027 Framework Programme (European Union) through the SuperMeQ project (Grant Agreement number 101080143), the European Research Council under Grant No. 101087847 (ERC Consolidator SuperQLev), and the Knut and Alice Wallenberg (KAW) Foundation through a Wallenberg Academy Scholar.
\end{acknowledgements}

\appendix

\section{Magnetic force and torque}\label{sec:eom}

The net force on the particle can be calculated by integrating the Lorentz force density over the particle volume,
\begin{align}
    {\bf F} = &\; \int_V d^3{\bf r} \,(\nabla \times {\bf M}) \times {\bf B}_{\rm ext} \nonumber\\
     = &\; \int_V d^3{\bf r}  \,\left[-\nabla ({\bf B}_{\rm ext} \cdot {\bf M}) + {\bf M} \times (\nabla \times {\bf B}_{\rm ext}) \right. \nonumber \\
      & \left. + ({\bf B}_{\rm ext}\cdot \nabla) {\bf M} + ({\bf M}\cdot \nabla) {\bf B}_{\rm ext} \right ].
\end{align}
Here, we dropped the dependence of all vector fields on ${\bf r}$. The first term vanishes by Gauss' theorem and the second term vanishes since $\nabla \times {\bf B}_{\rm ext} = 0$. The third term also vanishes as can be seen by rewriting $({\bf B}_{\rm ext} \cdot \nabla) {\bf M} = \nabla \cdot ({\bf B}_{\rm ext} \otimes {\bf M}) - (\nabla \cdot {\bf B}_{\rm ext}){\bf M}$, where the first term vanishes when integrated over due to Gauss' theorem and the second term vanishes due to Maxwell's equations. We thus obtain
\begin{equation}
    {\bf F} = \int_V d^3{\bf r} \,({\bf M}\cdot\nabla){\bf B}_{\rm ext} = {\rm C} {\bf m},
\end{equation}
where the last equality holds for quadrupole fields with constant quadrupole tensor $\nabla \otimes {\bf B}_{\rm ext} = {\rm C}$ and we used the definition of the magnetic dipole moment \eqref{eq:magdipdef}. Note that the dipole moment ${\bf m}$ is evaluated at the center-of-mass position ${\bf R}$, which enters through the integration volume $V = V({\bf R},\Omega)$.

The net torque for particle rotations around the center of mass ${\bf R}$ can be calculated in a similar fashion,
\begin{align}\label{eq:torquederiv}
{\bf N} = &\; \int_V d^3{\bf r}\, ({\bf r}-{\bf R}) \times [(\nabla \times {\bf M})\times {\bf B}_{\rm ext}]\nonumber \\
 = &\; \int_V d^3{\bf r}\, [({\bf r} - {\bf R}) \cdot {\bf B}_{\rm ext}] (\nabla \times {\bf M})\nonumber \\
 & - \int_V d^3{\bf r}\,[(\nabla \times {\bf M}) \cdot ({\bf r} - {\bf R}) ] {\bf B}_{\rm ext}.
\end{align}
The first integral on the right-hand side ${\bf N}_1$ can be evaluated by noting that
\begin{align}
     {\bf N}_1 =&\; \int_V d^3{\bf r}\, [({\bf r} - {\bf R}) \cdot {\bf B}_{\rm ext}] (\nabla \times {\bf M}) \nonumber \\
     = &\; \int_V d^3{\bf r}\, \nabla \times [({\bf r} - {\bf R}) \cdot {\bf B}_{\rm ext} {\bf M}] \nonumber\\
     & + \int_V d^3{\bf r} \,{\bf M} \times \nabla[({\bf r} - {\bf R}) \cdot {\bf B}].
\end{align}
The first term vanishes by Gauss' theorem and the second term can be rewritten by using that $\nabla \times {\bf B}_{\rm ext} = 0 $ and $\nabla \times ({\bf r} - {\bf R}) = 0$, yielding
\begin{align}
    {\bf N}_1 =&\; \int_V d^3{\bf r} \,{\bf M} \times \left \{[({\bf r} - {\bf R}) \cdot \nabla] {\bf B}_{\rm ext}\right.\nonumber\\
    &\hspace{1cm}\left. + ({\bf B}_{\rm ext} \cdot \nabla) ({\bf r} - {\bf R})\right \}
\end{align}
To proceed further we use that $\nabla \otimes {\bf B}_{\rm ext} = {\rm C}$ is homogeneous and that $\nabla \otimes ({\bf r} - {\bf R}) = \mathds{1}$. In addition, for quadrupole fields ${\bf B}_{\rm ext}({\bf r}) = {\bf B}_{\rm ext}({\bf R}) + {\rm C}({\bf r} - {\bf R})$ so that
\begin{equation}
    {\bf N}_1 = {\bf m}\times {\bf B}_{\rm ext}({\bf R}) + 2\int_V d^3{\bf r}\, {\bf M} \times {\rm C}({\bf r} - {\bf R}).
\end{equation}
Likewise, we can compute the second integral ${\bf N}_2$ on the right-hand side of Eq.~\eqref{eq:torquederiv},
\begin{align}
{\bf N}_2 = &\; - \int_V d^3{\bf r}\,[(\nabla \times {\bf M}) \cdot ({\bf r} - {\bf R}) ] \, {\bf B}_{\rm ext}\nonumber\\
= &\; -\int_V d^3{\bf r} \,  {\bf B}_{\rm ext}\,\nabla\cdot[{\bf M}\times ({\bf r} - {\bf R})] ,
\end{align}
where we used that $\nabla \times ({\bf r} -{\bf R})= 0$. This integral can be further evaluated by noting that
\begin{align}
    {\bf N}_2 = &\; -\int_Vd^3{\bf r} \,\nabla \cdot \left [ {\bf M}\times ({\bf r} - {\bf R}) \otimes {\bf B}_{\rm ext} \right ] \nonumber \\
     & + \int_Vd^3{\bf r} \,[{\bf M} \times ({\bf r} - {\bf R}) ]\cdot \nabla \,{\bf B}_{\rm ext} \nonumber\\
     = &\; {\rm C} \int_V d^3{\bf r} \,[{\bf M} \times ({\bf r} - {\bf R})],
\end{align}
where we used Gauss' theorem, again. Putting everything together and using the spectral representation of the quadrupole moment of the trapping field,
\begin{equation}
    {\rm C} = \sum_{i = 1}^3 C_i {\bf e}_i \otimes {\bf e}_i,
\end{equation}
with $\sum_i C_i = 0$, one obtains the torque
\begin{equation}
    {\bf N} = {\bf m} \times {\bf B}_{\rm ext} + \sum_{i = 1}^3 C_i ({\rm Q}{\bf e}_i) \times {\bf e}_i.
\end{equation}

\section{Definition of magnetic moments} \label{sec:momentsgen}

We start by showing how the magnetic dipole vector ${\bf m}$ and the quadruopole tensor ${\rm Q}$ can be calculated from the magnetization field ${\bf M}({\bf r})$. The magnetic field ${\bf B}_{\rm mag} = \nabla \times {\bf A}_{\rm mag}$ sourced by the magnetization field ${\bf M}$ follows from the vector potential
\begin{equation}
{\bf A}_{\rm mag}({\bf r}) = \frac{\mu_0}{4\pi} \int d^3{\bf s}\, \frac{\nabla_{\bf s} \times {\bf M}({\bf s})}{|{\bf r} - {\bf s}|}.
\end{equation}
Using that the magnetization field vanishes outside the particle, this expression can be rewritten by using Gauss' theorem as
\begin{equation}
    {\bf A}_{\rm mag}({\bf r}) = \frac{\mu_0}{4\pi} \int d^3{\bf s}\,  {\bf M}({\bf s}) \times \frac{{\bf r} - {\bf s}}{|{\bf r} - {\bf s}|^3}.
\end{equation}
For localized magnetization fields, the vector potential outside the body can be expanded in its multipole moments by using that
\begin{equation}\label{eq:expansion}
    \frac{{\bf r} - {\bf s}}{|{\bf r} - {\bf s}|^3} \approx  \frac{{\bf r} - {\bf R}}{|{\bf r} - {\bf R}|^3} +4 \pi {\rm G}({\bf r} - {\bf R}) ({\bf s}-{\bf R}).  
\end{equation}
where ${\bf R}$ is the particle center-of-mass position and we used that the Green tensor 
\begin{equation}\label{eq:greentensor}
{\rm G}({\bf r}) = \frac{1}{4\pi} \nabla \otimes \nabla \frac{1}{|{\bf r}|} = \frac{3 {\bf r} \otimes {\bf r} - |{\bf r}|^2 \mathds{1}_3}{4\pi |{\bf r}|^5} ,
\end{equation}
for $|{\bf r}| \neq 0$. Inserting this expansion, yields the vector potential
\begin{align}\label{eq:vecpotderiv}
    {\bf A}_{\rm mag}({\bf r}) = &\; \frac{\mu_0}{4\pi}\frac{{\bf m} \times ({\bf r} - {\bf R})}{|{\bf r} - {\bf R}|^3} + {\bf A}_{\rm qu}({\bf r}),
\end{align}
where we defined the magnetic dipole moment
\begin{equation}
    {\bf m} = \int_V d^3 {\bf r} \, {\bf M}({\bf r}).
\end{equation}
The second term in Eq.~\eqref{eq:vecpotderiv} is the quadrupole vector potential
\begin{align}\label{eq:aquad}
    {\bf A}_{\rm qu}({\bf r}) = &\;  \mu_0 \int d^3{\bf s}\,  {\bf M}({\bf s}) \times {\rm G}({\bf r} - {\bf R}) ({\bf s} - {\bf R}) \nonumber \\
    = &\; \frac{\mu_0}{8\pi} \nabla \cdot {\rm Q}\times \nabla \frac{1}{|{\bf r} - {\bf R}|} \\
    & - \frac{\mu_0}{8\pi} \nabla \int_V d^3{\bf s}\, [{\bf M}({\bf s})\times ({\bf s} - {\bf R})] \cdot \nabla\frac{1}{|{\bf r} - {\bf R}|}, \nonumber
\end{align}
where the second equality follows from using Eq.~\eqref{eq:greentensor} and from defining the magnetic quadrupole tensor
\begin{align}
    {\rm Q} = &\; \int_V d^3{\bf r}\, \left [ \vphantom{\frac{2}{3}} 3 {\bf M}({\bf r}) \otimes ({\bf r} - {\bf R}) + 3 ({\bf r} - {\bf R}) \otimes {\bf M}({\bf r})\right.\nonumber\\
    & \left. - 2 {\bf M}({\bf r}) \cdot ({\bf r} - {\bf R}) \mathds{1}\right ].
\end{align}
Note that the last term in Eq.~\eqref{eq:aquad} is a gradient with respect to ${\bf r}$ and thus does not contribute to the quadrupole field ${\bf B}_{\rm qu} = \nabla \times {\bf A}_{\rm qu}$. Expressions for higher order moments can be derived in a similar fashion by taking higher order terms in the expansion \eqref{eq:expansion} into account.

\section{Einstein-de Haas and Barnett effects}\label{sec:edhbarnett}

The Einstein-de Haas and Barnett effects describe the contribution of a body's magnetization to its total angular momentum and vice versa. We divide the body into small volumes $V_m$ and denote by ${\bf m}_{\alpha k m}$ the magnetic moment of the $k$-th microscopic dipole of type $\alpha$ in $V_m$. The index $\alpha$ labels distinct magnetic contributions such as orbital motion of electrons, electronic spins, or nuclear spins \cite{white1983quantum}. Each magnetic moment also carries angular momentum ${\bf L}_{\alpha k m} = {\bf m}_{\alpha k m}/\gamma_\alpha$ as quantified by the material-specific gyromagnetic ratio $\gamma_\alpha$. The magnetization ${\bf M}({\bf r})$ quantifies the local density of magnetic moments,
\begin{align}
    {\bf M}({\bf r}) = &\;  \sum_\alpha \sum_m \frac{\Theta({\bf r} \in V_m)}{V_m} \sum_k {\bf m}_{\alpha k m}  \nonumber \\
    = &\; \sum_\alpha \gamma_\alpha \sum_m \frac{\Theta({\bf r} \in V_m)}{V_m} \sum_k {\bf L}_{\alpha k m}  ,
\end{align}
where $\Theta({\bf r} \in V_m)$ is unity if ${\bf r}\in V_m$ and zero otherwise. Consequently, the total magnetic dipole moment of the body
\begin{equation}
{\bf m} = \int_V d^3{\bf r} \, {\bf M}({\bf r}) =   \sum_\alpha \gamma_\alpha \sum_{mk} {\bf L}_{\alpha k m}
\end{equation}
and the total internal angular momentum
\begin{equation}
    {\bf L}_{\rm int} = \sum_{\alpha k m}{\bf L}_{\alpha k m}
\end{equation}
are not independent. Their connection implies that any change in ${\bf m}$ is accompanied by a change of ${\bf L}_{\rm int}$ which in turn, in the absence of external torques, must be compensated by the mechanical rotation of the body due to the conservation of total angular momentum. This conversion from magnetization into mechanical rotation is known as the Einstein-de Haas effect.

To see how a body's mechanical rotation affects its magnetization, we consider the conservative dynamics of the microscopic dipole moments 
\begin{equation}
    \frac{d}{dt} {\bf m}_{\alpha k m} = \gamma_\alpha {\bf m}_{\alpha k m} \times {\bf B}_{km}.
\end{equation}
They evolve according to the magnetic field ${\bf B}_{km}$ at the dipole location. This field can be divided into two contributions, ${\bf B}_{km} = {\bf B}_{m} + {\bf B}_{km}^{\rm dem}$, where ${\bf B}_{m}$ denotes the sum of all fields originating outside of $V_m$ (approximated as homogeneous in $V_m$) and ${\bf B}_{km}^{\rm dem}$ denotes the effective demagnetization field that describes the magnetic and exchange interaction with all other dipoles in $V_m$. Solving the coupled equations in $V_m$ perturbatively for small ${\bf B}_m$ and in the presence of dissipation, as for instance described by the Landau-Lifshitz-Gilbert equation, yields  effective models for the local static magnetic susceptibility tensor $\chi_{r,\alpha}({\bf r})$ \cite{kittel1948theory,polder1949viii}. This susceptibility relates the local magnetic field to the induced magnetization field ${\bf M}_\alpha({\bf r})$ of magnetic species $\alpha$, so that for isotropic materials ${\bf M}_\alpha({\bf r}) = \chi_{r,\alpha}({\bf r}) {\bf B}({\bf r})/\mu_0$ and ${\bf M}({\bf r}) = \sum_\alpha {\bf M}_\alpha({\bf r})$. The relation between local field and magnetization field for a single magnetic type is the starting point for the calculation in App.~\ref{sec:inducedmagn}, which shows how the magnetic interaction between different volumes $V_m$ can be included.

To see the impact of the body's rotation, we rewrite the equations of motion in terms of body-fixed vectors, indicated by a tilde, i.e.\ ${\bf m}_{\alpha k m} = {\rm R}(\Omega) \widetilde{\bf m}_{\alpha k m}$ and likewise for ${\bf B}_{m}$ and ${\bf B}_{km}^{\rm dem}$. Using that the angular velocity $\boldsymbol{\omega}$ is related to the change of the rotation tensor via $d{\rm R}/dt = \boldsymbol{\omega}\times {\rm R}$ yields
\begin{equation}
\frac{d}{dt} \widetilde{\bf m}_{\alpha k m}  = \gamma_\alpha \widetilde{\bf m}_{\alpha k m}\times \left [ \widetilde{\bf B}_{m} + \frac{\boldsymbol{\omega}}{\gamma_\alpha}   + \widetilde{\bf B}_{k m}^{\rm dem} \right ].
\end{equation}
As a consequence, the induced magnetization field is
\begin{equation}\label{eq:constitutiverotating}
    {\bf M}({\bf r}) = \frac{1}{\mu_0}\sum_\alpha \chi_{r,\alpha}({\bf r})\left [ {\bf B}({\bf r}) + \frac{\boldsymbol{\omega}}{\gamma_\alpha} \right ],
\end{equation}
implying that mechanical rotation can act as a synthetic magnetic field and induce a magnetization even in the absence of genuine magnetic fields. This relation between the mechanical rotation and an object's magnetization is known as the Barnett effect.

If only a single magnetic species contributes to the local magnetization, the sum over $\alpha$ drops from the above relations. In the case of superconducting particles, the induced magnetization field is caused by the orbital motion of Cooper pairs, so that
\begin{equation}
    \gamma_{\rm sc} = -\frac{e_0}{2m_e} = -\gamma_0 \approx -0.88 \times 10^{11}~{\rm T}^{-1} {\rm s}^{-1}.
\end{equation}
and $\chi_{\rm sc} = -1$. The rotation-induced magnetic moment ${\bf m}_L$, known as the London moment, of a superconducting sphere with volume $V_0$ and angular velocity $\boldsymbol{\omega}$ follows from the induced dipole moment \eqref{eq:magdip} with synthetic magnetic field $-\boldsymbol{\omega}/\gamma_0$ as
\begin{equation}
    {\bf m}_{L} = \frac{3 V_0}{2 \mu_0 \gamma_0} \boldsymbol{\omega},
\end{equation}
consistent with for instance Ref.~\cite{buchman1994applications}.

\end{document}